\documentclass[twocolumn]{aastex63}

\usepackage{mathtools}

\newcommand{\ra}[4]{$#1\overset{\mathrm h}{\phd}#2\overset{\mathrm m}{\phd}#3\overset{\mathrm s}{.}#4$}
\newcommand{\dec}[4]{$#1\overset{\circ}{\phd}#2\overset{\prime}{\phd}#3\overset{\prime\prime}{.}#4$}

\usepackage[multiple]{footmisc}

\received{}
\revised{}
\accepted{}

\shorttitle{CSM Constraints in Type Ia Supernovae}
\shortauthors{Sand et al.}

\graphicspath{{./}{figures/}}

\begin{document}

\title{Circumstellar Medium Constraints on the Environment of Two Nearby Type Ia Supernovae: SN~2017cbv and SN~2020nlb}

\correspondingauthor{D. J. Sand}
\email{dsand@arizona.edu}

\newcommand{\Thacher}{\affiliation{Thacher Observatory, Thacher School, 5025 Thacher Rd. Ojai, CA 93023, USA}}
\newcommand{\UA}{\affiliation{Steward Observatory, University of Arizona, 933 North Cherry Avenue, Tucson, AZ 85721-0065, USA}}
\newcommand{\NU}{\affiliation{Center for Interdisciplinary Exploration and Research in Astrophysics and Department of Physics and Astronomy, \\ Northwestern University, 2145 Sheridan Road, Evanston, IL 60208-3112, USA}}
\newcommand{\UCDavis}{\affiliation{Department of Physics, University of California, 1 Shields Avenue, Davis, CA 95616-5270, USA}}
\newcommand{\Padova}{\affiliation{Department of Physics and Astronomy Galileo Galilei, University of Padova, Vicolo dell'Osservatorio, 3, I-35122 Padova, Italy}}
\newcommand{\INAF}{\affiliation{INAF Osservatorio Astronomico di Padova, Vicolo dell'Osservatorio 5, I-35122 Padova, Italy}}
\newcommand{\INAFbol}{\affiliation{INAF - Osservatorio di Astrofisica e Scienza dello Spazio - Via Piero Gobetti 93/3, I-40129 Bologna, Italy}}
\newcommand{\LPL}{\affiliation{Lunar and Planetary Lab, Department of Planetary Sciences, University of Arizona, Tucson, AZ 85721, USA}}
\newcommand{\NOAO}{\affiliation{National Optical Astronomy Observatory, 950 North Cherry Avenue, Tucson, AZ 85719, USA}}
\newcommand{\PATAU}{\affiliation{The School of Physics and Astronomy, Tel Aviv University, Tel Aviv 69978, Israel}}
\newcommand{\TTU}{\affiliation{Department of Physics and Astronomy, Texas Tech University, Box 1051, Lubbock, TX 79409-1051, USA}}
\newcommand{\OSU}{\affiliation{Department  of  Astronomy,  The  Ohio  State University,  140  W.  18th  Ave.,  Columbus,  OH43210, USA}}
\newcommand{\LBT}{\affiliation{Large Binocular Telescope Observatory, 933 North Cherry Avenue, Tucson, AZ, USA}}
\newcommand{\RMC}{\affiliation{Department of Physics and Space Science Royal Military College of Canada P.O. Box 17000, Station Forces Kingston, ON K7K 7B4, Canada}}
\newcommand{\ASU}{\affiliation{School of Earth and Space Exploration, Arizona State University, Tempe, AZ 85287, USA}}
\newcommand{\MMT}{\affiliation{MMT Observatory, PO Box 210065, University of Arizona, Tucson, AZ 85721-0065, USA}}
\newcommand{\NAU}{\affiliation{Department of Physics and Astronomy, Northern Arizona University, P.O. Box 6010, Flagstaff, AZ 86011, USA}}
\newcommand{\UAOptSci}{\affiliation{College of Optical Sciences, University of Arizona, 1630 E University Blvd, Tucson, AZ 85719, USA}}
\newcommand{\UNC}{\affiliation{Department of Physics and Astronomy, University of North Carolina at Chapel Hill, Chapel Hill, NC 27599, USA}}
\newcommand{\MSU}{\affiliation{Center for Data Intensive and Time Domain Astronomy, Department  of  Physics  and  Astronomy,  Michigan  State  University,East Lansing, MI 48824, USA}}
\newcommand{\UCSC}{\affiliation{Department of Astronomy and Astrophysics, University of California, Santa Cruz, CA 95064, USA}}
\newcommand{\STScI}{\affiliation{Space Telescope Science Institute, 3700 San Martin Drive, Baltimore, MD 21218, USA}}
\newcommand{\Brandeis}{\affiliation{Department of Physics, Brandeis University, Waltham, MA 02453, USA}}
\newcommand{\LCO}{\affiliation{Las Cumbres Observatory, 6740 Cortona Drive, Suite 102, Goleta, CA 93117-5575, USA}}
\newcommand{\UToronto}{\affiliation{Department of Astronomy and Astrophysics, University of Toronto, 50 St. George Street, Toronto, Ontario, M5S 3H4 Canada}}
\newcommand{\NotreDame}{\affiliation{Department of Physics, University of Notre Dame, Notre Dame, IN 46556, USA}}
\newcommand{\UMN}{\affiliation{College of Science \& Engineering, Minnesota Institute for Astrophysics, University of Minnesota, 115 Union St. SE, Minneapolis, MN 55455, USA}}
\newcommand{\UT}{\affiliation{Department of Astronomy, University of Texas at Austin, Austin, TX 78712, USA}}
\newcommand{\JHU}{\affiliation{The Johns Hopkins University, Baltimore, MD 21218, USA}}
\newcommand{\VAT}{\affiliation{Vatican Observatory, 00120 Citt\`{a} del Vaticano, Vatican City State  }}
\newcommand{\HF}{\affiliation{Hubble Fellow}}
\newcommand{\Carnegie}{\affiliation{The Observatories of the Carnegie Institution for Science, 813 Santa Barbara St., Pasadena, CA 91101, USA}}

\author[0000-0003-4102-380X]{D.~J. Sand}
\UA

\author[0000-0002-4781-7291]{S. K. Sarbadhicary}
\affil{Department of Physics and Astronomy, Michigan State University, East Lansing, MI 48824, USA}

\author[0000-0002-7472-1279]{C. Pellegrino}
\LCO
\affil{Department of Physics, University of California, Santa Barbara, CA 93106-9530, USA}

\author{K. Misra}
\affiliation{Aryabhatta Research Institute of Observational Sciences, Manora Peak, Nainital 263 001, India}

\author[0000-0001-6191-7160]{R. Dastidar}
\affiliation{Millennium Institute of Astrophysics, Nuncio Monsenor Sotero Sanz 100, Providencia, Santiago, Chile}

\author[0000-0001-6272-5507]{P. J. Brown}
\affil{Department of Physics and Astronomy, Texas A\&M University, 4242 TAMU, College Station, TX 77843, USA}
\affil{George P. and Cynthia Woods Mitchell Institute for Fundamental Physics \& Astronomy}

\author{K.~Itagaki }
\affil{Itagaki Astronomical Observatory, Yamagata 990-2492, Japan}

\author[0000-0001-8818-0795]{S.~Valenti}
\affiliation{Department of Physics and Astronomy, University of California, 1 Shields Avenue, Davis, CA 95616-5270, USA}

\author[0000-0002-9486-818X]{Jonathan J. Swift}
\Thacher

\author[0000-0003-0123-0062]{J.~E. Andrews}
\affiliation{Gemini Observatory/NSF’s NOIRLab, 670 N. A’ohoku Place, Hilo, Hawai’i, 96720, USA}




\author[0000-0002-4924-444X]{K. A. Bostroem}
\affiliation{Department of Physics and Astronomy, University of California, 1 Shields Avenue, Davis, CA 95616-5270, USA}

\author[0000-0003-0035-6659]{J. Burke}
\affiliation{Department of Physics, University of California, Santa Barbara, CA 93106-9530, USA}
\LCO

\author[0000-0002-8400-3705]{L. Chomiuk}
\affil{Department of Physics and Astronomy, Michigan State University, East Lansing, MI 48824, USA}

\author[0000-0001-8818-0795]{Y.~Dong}
\affiliation{Department of Physics and Astronomy, University of California, 1 Shields Avenue, Davis, CA 95616-5270, USA}

\author[0000-0002-1296-6887]{L.~Galbany}
\affiliation{Institute of Space Sciences (ICE, CSIC), Campus UAB, Carrer de Can Magrans, s/n, E-08193 Barcelona, Spain}

\author[0000-0002-9154-3136]{M.~L.~Graham}
\affiliation{DiRAC Institute, Department of Astronomy, University of Washington, Box 351580, U.W., Seattle, WA 98195, USA}

\author[0000-0002-1125-9187]{D. Hiramatsu}
\affiliation{Department of Physics, University of California, Santa Barbara, CA 93106-9530, USA}
\LCO

\author[0000-0003-4253-656X]{D. A. Howell}
\LCO

\author[0000-0003-1039-2928]{E. Y. Hsiao}
\affil{Department of Physics, Florida State University, 77 Chieftan Way, Tallahassee, FL 32306, USA}

\author[0000-0003-0549-3281]{D. Janzen}
\affiliation{Department of Physics and Engineering Physics, University of Saskatchewan, 116 Science Pl, Saskatoon, SK S7N 5E2, Canada}

\author[0000-0001-5754-4007]{J. E. Jencson}
\affil{Steward Observatory, University of Arizona, 933 North Cherry Avenue, Tucson, AZ 85721-0065, USA}

\author[0000-0001-9589-3793]{M.~J. Lundquist}
\UA

\author[0000-0001-5807-7893]{C. McCully}
\affiliation{Department of Physics, University of California, Santa Barbara, CA 93106-9530, USA}
\LCO

\author[0000-0002-5060-3673]{D. Reichart}
\UNC

\author[0000-0001-5510-2424]{Nathan Smith}
\affil{Steward Observatory, University of Arizona, 933 North Cherry Avenue, Tucson, AZ 85721-0065, USA}

\author[0000-0002-1094-3817] {Lingzhi Wang}
\affiliation{Chinese Academy of Sciences South America Center for Astronomy, National Astronomical Observatories, CAS, Beijing 100101, People's Republic of China}
\affiliation{CAS Key Laboratory of Optical Astronomy, National Astronomical Observatories, Chinese Academy of Sciences, Beijing 100101, People's Republic of China}


\author[0000-0003-2732-4956]{S.~Wyatt}
\UA

\begin{abstract}
We present deep {\it Chandra} X-ray observations of two nearby Type Ia supernovae, SN~2017cbv and SN~2020nlb, which reveal no X-ray emission down to a luminosity $L_X$$\lesssim$5.3$\times$10$^{37}$ and $\lesssim$5.4$\times$10$^{37}$ erg s$^{-1}$ (0.3--10 keV), respectively, at $\sim$16--18 days after the explosion.  With these limits, we constrain the pre-explosion mass-loss rate of the progenitor system to be $\dot{M}$$<$7.2$\times$10$^{-9}$ and $<$9.7$\times$10$^{-9}$ M$_{\odot}$ yr$^{-1}$ for each (at a wind velocity $v_w$=100 km s$^{-1}$ and a radius of $R$$\approx$10$^{16}$ cm), assuming any X-ray emission would originate from inverse Compton emission from optical photons up-scattered by the supernova shock.  If the supernova environment was a constant density medium, we find a number density limit of n$_{CSM}$$<$36 and $<$65 cm$^{-3}$, respectively. These X-ray limits rule out all plausible symbiotic progenitor systems, as well as large swathes of parameter space associated with the single degenerate scenario, such as mass loss at the outer Lagrange point and accretion winds. We also present late-time optical spectroscopy of SN~2020nlb, and set strong limits on any swept up hydrogen ($L_{H\alpha}$$<$2.7$\times$10$^{37}$ ergs s$^{-1}$) and helium ($L_{He, \lambda 6678}$$<$2.7$\times$10$^{37}$ ergs s$^{-1}$) from a nondegenerate companion, corresponding to $M_{H}$$\lesssim$0.7--2$\times$10$^{-3}$ M$_{\odot}$ and $M_{He}$$\lesssim$4$\times$10$^{-3}$ M$_{\odot}$. 
Radio observations of SN~2020nlb at 14.6 days after explosion also yield a non-detection, ruling out most plausible symbiotic progenitor systems.  While we have doubled the sample of normal type Ia supernovae with deep X-ray limits, more observations are needed to sample the full range of luminosities and sub-types of these explosions, and set statistical constraints on their circumbinary environments.
\end{abstract}


\section{Introduction} \label{sec:intro}

Despite their critical use for cosmology, the exact progenitor systems and explosion mechanisms for Type Ia supernovae (SNe~Ia) are still being determined \citep[e.g.,][for a recent review]{Jha19}.  There are two general categories of SN~Ia progenitors which can plausibly explain how the carbon-oxygen white dwarf accretes the necessary mass to cause an explosion: the single degenerate (SD) and double degenerate (DD) scenarios.  In the DD scenario a second degenerate companion (i.e. another white dwarf) is in the binary system \citep{Iben84,Webbink84}, while in the SD scenario there is a nondegenerate companion star \citep{Whelan73}.  Within these two broad categories, the exact triggering mechanism for the thermonuclear explosion is a topic of current research.

There are several observational techniques that have been developed to shed light on the progenitor system, although we only mention a few here.   For instance, models predict that the very early light curves of SNe Ia may exhibit a `blue bump' in the UV-optical in the days after explosion due to the ejecta shocking the nondegenerate companion \citep[][]{Kasen10}. A similar signature has been observed in a handful of instances \citep[e.g.][and strongly constrained in other instances, e.g. \citealt{Hayden10,Bianco11,Ganes11,Brown12,Olling15}]{Cao15,Marion16,Hosseinzadeh17,Miller18,Shappee19,Dimitriadis19_18oh,Miller20,Tucker2019yvq,Burke21}, although the interpretation of these early light curves are still a matter of debate.  Meanwhile, another prediction of the SD scenario is that material stripped from the companion star is swept up by the SN Ia ejecta, detectable as a narrow emission line of hydrogen or helium at late times \citep[most recently,][]{Boty18,Dessart20}.  While most searches have only led to limits on the amount of stripped hydrogen and/or helium \citep{Mattila05,Leonard07,Shappee13,Lundqvist13,Lundqvist15,Maguire16,Graham17,Shappee18,Holmbo18,Tucker19,Dimitriadis19,Sand16,Sand18_2017cbv,Sand19,Tucker20}, three recent detections from fast-declining SNe Ia \citep[][]{Kollmeier19,Vallely19_18tb,Prieto20,Elias21} indicate that a broader search for such features may prove fruitful. 
Presumably, if an early light curve bump was due to interaction with a nondegenerate companion, then at late times a hydrogen or helium emission line should be visible.  
Such important cross-checks between early light curve and nebular signatures of the progenitor are essential, as has recently been carried out for SN~2017cbv (early light curve: \citealt{Hosseinzadeh17}; nebular spectra: \citealt{Sand18_2017cbv}) and SN~2018oh (early light curve: \citealt{Dimitriadis19_18oh,Shappee19,Li19}; nebular spectra: \citealt{Dimitriadis19,Tucker19}).  In these recent examples, even though the early light curves may point to a single degenerate progenitor, the nebular spectra do not necessarily corroborate this picture, suggesting that further progenitor probes are necessary to understand these systems.

One powerful probe of the immediate SN~Ia environment, and thus the progenitor system, utilizes X-ray observations around the time of maximum light \citep{Margutti12,Horesh12,Margutti14,Russell12,Shappee18,Shappee19,Stauffer21}. X-ray emission in the weeks after a SN Ia explosion originates from the upscattering of optical photons off relativistic particles accelerated at the SN shock  \citep[i.e. inverse Compton emission;][]{Chevalier06,Margutti12}.  This circumstellar medium was in turn shaped by the mass loss of the progenitor star system leading up to the explosion.  Broadly speaking, in the SD scenario the white dwarf accretes material from a nondegenerate stellar companion, either through direct Roche lobe overflow \citep[RLOF;][]{Nomoto82} or from a wind from the secondary star (the `symbiotic channel'; e.g. \citealt{Patat11}) -- it thus is expected to have residual CSM in its immediate environment either from the donor star wind or from non-conservative mass loss from the RLOF transfer.  By contrast, a standard DD scenario involving two white dwarfs inspiraling due to angular momentum loss should have a `cleaner' CSM environment.

Deep {\it Chandra} X-ray limits have been reported for only two normal SNe Ia thus far -- the nearby and well studied SN~2011fe \citep{Margutti12} and SN~2014J \citep{Margutti14}.  Both resulted in strong X-ray upper limits which ruled out a symbiotic giant star companion and large portions of parameter space associated with accretion winds and Lagrangian losses from a main sequence or subgiant companion.  Despite these strong results, deep X-ray data for two SNe Ia is insufficient to draw wider conclusions about the SN Ia population, and ultimately a statistical data set of deep X-ray limits on SNe Ia is needed (along with other tracers) to constrain SN~Ia progenitor properties.  

Here we present {\it Chandra} X-ray data of two nearby type Ia SNe -- SN~2017cbv and SN~2020nlb  -- to constrain the circumstellar environment associated with the progenitor system, doubling the sample with deep X-ray limits.  The study of SN~2017cbv is particularly interesting given the `blue bump' observed in its early light curve, which may be due to interaction with a nondegenerate companion.  Meanwhile, we couple deep X-ray data of SN~2020nlb with radio data from the Very Large Array (VLA) and nebular spectroscopy to constrain any hydrogen or helium emission -- these complementary probes are essential for narrowing in on the SN Ia progenitor.  We end this paper with our X-ray derived CSM constraints and a discussion of the donor star and allowed progenitor configuration for SN~2017cbv and SN~2020nlb.

\section{Background}\label{sec:back}

\subsection{SN~2017cbv}

SN~2017cbv (RA \ra{14}{32}{34}{38} \& Dec \dec{-44}{08}{03}{1} J2000) was discovered in the outskirts of the nearby galaxy NGC~5643 on 2017 March 10 UT (MJD 57822.14) at a magnitude of $r$$\approx$16 by the Distance Less Than 40 Mpc survey \citep[DLT40; ][]{Tartaglia18}.  Within hours of discovery, the transient was classified as a very young SN Ia \citep{SN17cbv_classify}, and an intense multiwavelength follow-up campaign was begun.  The early light curve of SN~2017cbv displayed a clear, blue excess, lasting $\sim$3 days \citep{Hosseinzadeh17}.  Such excess emission is predicted in the single degenerate scenario, when the SN ejecta shock a nondegenerate companion star \citep{Kasen10}.  While the observed optical light curve of SN~2017cbv is well fit to this model, it overpredicts the observed flux in the ultraviolet bands \citep[possibly due to line blanketing or other physics not currently accounted for in the models; see,][]{Hosseinzadeh17}. Nebular phase spectroscopy did not detect any narrow hydrogen or helium features \citep{Sand18_2017cbv}, as might be expected in the single degenerate scenario \citep[e.g.][]{Boty18,Dessart20}, although further modeling is needed.  The uncertain origin for the early blue excess emission makes SN~2017cbv an excellent target for a deep X-ray search for CSM interaction, as a complementary probe of the progenitor system. 

We adopt an explosion epoch of MJD 57821.0, as found by \citet{Hosseinzadeh17} based on the early Si II velocity evolution \citep[using the technique described in ][]{Piro14}.
We also adopt a distance of $\mu$=30.58 mag ($D$$\approx$13.1 Mpc), the value used by \citet{Wang20} for presenting their SN~2017cbv bolometric light curve, which we adopt for our analysis as well.  This distance is in agreement with other recent light curve analyses of SN~2017cbv \citep[e.g.][]{Sand18_2017cbv,Burns20}, and a tip of the red giant branch distance to SN~2017cbv's host galaxy, NGC~5643 \citep{Hoyt21}.  We also use the epoch of $B$-band maximum from \citealt{Wang20} for reference, MJD 57840.87.

The Galactic neutral hydrogen column density in the direction of SN~2017cbv is N$_{\rm H}$ = 8.014$\times$10$^{20}$ cm$^{-2}$ \citep{Kalberla05}.  The hydrogen column associated with the host galaxy is likely negligible, as no significant host galaxy extinction is evident \citep[see discussions in][]{Ferretti17,Sand18_2017cbv}; we therefore adopt N$_{\rm H}$ = 8.014$\times$10$^{20}$ cm$^{-2}$ as the total hydrogen column density to SN~2017cbv in this work.



\subsection{SN~2020nlb} \label{sec:2020nlb}

SN~2020nlb was discovered on 2020 June 25.25 UT (MJD 59025.25) by the ATLAS survey \citep{tonry18} with an {\it orange} magnitude of 17.44.  The last non-detection from ATLAS was two days earlier (2020 June 23.28). The field of SN~2020nlb was also monitored by the Itagaki Astronomical Observatory's 0.35-m telescope in Okayama, Japan who obtained a tighter nondetection epoch of 2020 June 24.57 (MJD 59024.57), with a limiting magnitude of $<$18.5 mag, before also detecting the SN on 2020 June 26.56 at 16.1 mag.  The unfiltered Itagaki photometry was extracted using \texttt{Astrometrica} \citep{Raab2012} and calibrated to the Fourth US Naval Observatory CCD Astrograph Catalog (UCAC4; \citealt{Zacharias2013}).  We adopt the midpoint between the Itagaki non-detection and the first detection by the ATLAS survey as the explosion epoch (MJD 59024.91). Our X-ray analysis is not sensitive to the exact epoch adopted.

SN~2020nlb (RA \ra{12}{25}{24}{18} \& Dec \dec{+18}{12}{12}{5} J2000) exploded in the halo of M85 \cite[$v_{\odot}$=729 km s$^{-1}$, $z_{\odot}$=0.002432;][]{Smith00}, an early type galaxy in the Virgo Cluster.  Many distance measurements to M85 have been made, but we will utilize the surface brightness fluctuation measurement from the ACS Virgo Cluster Survey, which found a distance modulus of $\mu$=31.26$\pm$0.05 mag \citep[$D$=17.9 Mpc;][]{Mei07}.  We adopt a Milky Way extinction of $E(B-V)$=0.026 mag based on the dust maps of \citet{Schlafly11}.



No results on SN~2020nlb have been published in the  literature, and so we present UV/optical photometry and a maximum-light spectrum to characterize this SN Ia.  We also present a nebular spectrum to constrain narrow hydrogen or helium emission from any companion or CSM interaction to complement the X-ray CSM constraints. A more comprehensive optical-infrared analysis of SN~2020nlb will be presented in a future work.  
The Galactic neutral hydrogen column density in the direction of SN~2020nlb is N$_{\rm H}$ = 2.49$\times$10$^{20}$ cm$^{-2}$ \citep{Kalberla05}.

\subsubsection{Light Curve}\label{sec:SN20nlb_lc}
We display a $UBVgri$ light curve of SN~2020nlb in the left panel of Figure~\ref{fig:lc_spec}, taken with the 0.4- and 1.0-m telescope network of Las Cumbres Observatory \citep{Brown13} as part of the Global Supernova Project \citep[e.g.][]{Szalai19}.  These data were reduced in a standard way using the \texttt{PyRAF}-based photometric pipeline \texttt{lcogtsnpipe} \citep{Valenti16}.   An additional, high cadence data set was obtained by the 0.7m Thacher Observatory \citep{Thacher} in the $g,r,i,z$ bands.  The data was reduced in a standard way, and the PSF photometry tool \texttt{DoPHOT} \citep{Schechter93} was used in conjunction with photometric calibration from Pan-STARRS DR1 \citep{PS1_data} to produce the final light curve. Since the SN was offset from the host galaxy, no image subtraction was performed,  and we expect host galaxy contamination to be minimal. 

Observations from the  Neil Gehrels {\it Swift} Observatory \citep[{\it Swift};][]{Gehrels04} Ultra-Violet Optical Telescope \citep[UVOT;][]{Roming05} were also obtained and reduced using the pipeline associated with the {\it Swift} Optical Ultraviolet Supernovae Archive \citep[SOUSA;][]{Brown_etal_2014_SOUSA} and the zeropoints of \citet{Breeveld10}.  The {\it Swift} data is also displayed in Figure~\ref{fig:lc_spec}.

We fit a fourth order polynomial to the $B$ and $V$-band light curve of SN~2020nlb around maximum light, resampling the data based on the photometric uncertainties over 1000 trials.  We find a peak $B$-band apparent magnitude of $B_{max}$=12.11$\pm$0.02 mag on MJD = 59041.8 $\pm$ 0.2 (UT 2020 July 11.8), which we adopt throughout this work.  The $V$-band light curve peaked at $V_{max}$=12.04$\pm$0.03 mag on MJD = 59043.9 $\pm$ 0.3, which is 2.1 d after the $B$-band peak.  The $B$-band decline rate $\Delta$m$_{15}$($B$) is measured to be 1.29$\pm$0.06 mag, based on the polynomial fit.  

One way to constrain host galaxy extinction is to measure the color at maximum light, $B_{max}$$-$$V_{max}$, which was parameterized as a function of the decline rate parameter, $\Delta$$m_{15}$$(B)$, in \citet{Phillips99}.  After applying a Milky Way extinction of $E(B-V)_{MW}$=0.026 mag, we find $B_{max}$$-$$V_{max}$=0.04$\pm$0.04 mag from the polynomial fits described in the previous paragraph.  This is close to the expectation from the \citet{Phillips99} relation ($B_{max}$-$V_{max}$=$-$0.05$\pm$0.03 mag), and so we consider host extinction to be minimal for this work.  Any underestimate of the extinction would lead to a corresponding underestimate of the luminosity of the supernova, which ultimately would yield slightly weaker constraints on the CSM from our X-ray limits.  Applying only Milky Way extinction to the observed $B_{max}$, and using a distance of 17.9 Mpc, yields a peak absolute $B$-band magnitude of $M_{B}$=$-$19.25 mag, which is in line with SNe Ia with similar decline rates \citep{Blondin12,Folatelli13}.


\subsubsection{Spectroscopy}\label{sec:sn20nlb_spec}

\begin{figure*}
\centering
\includegraphics[width=8.9cm]{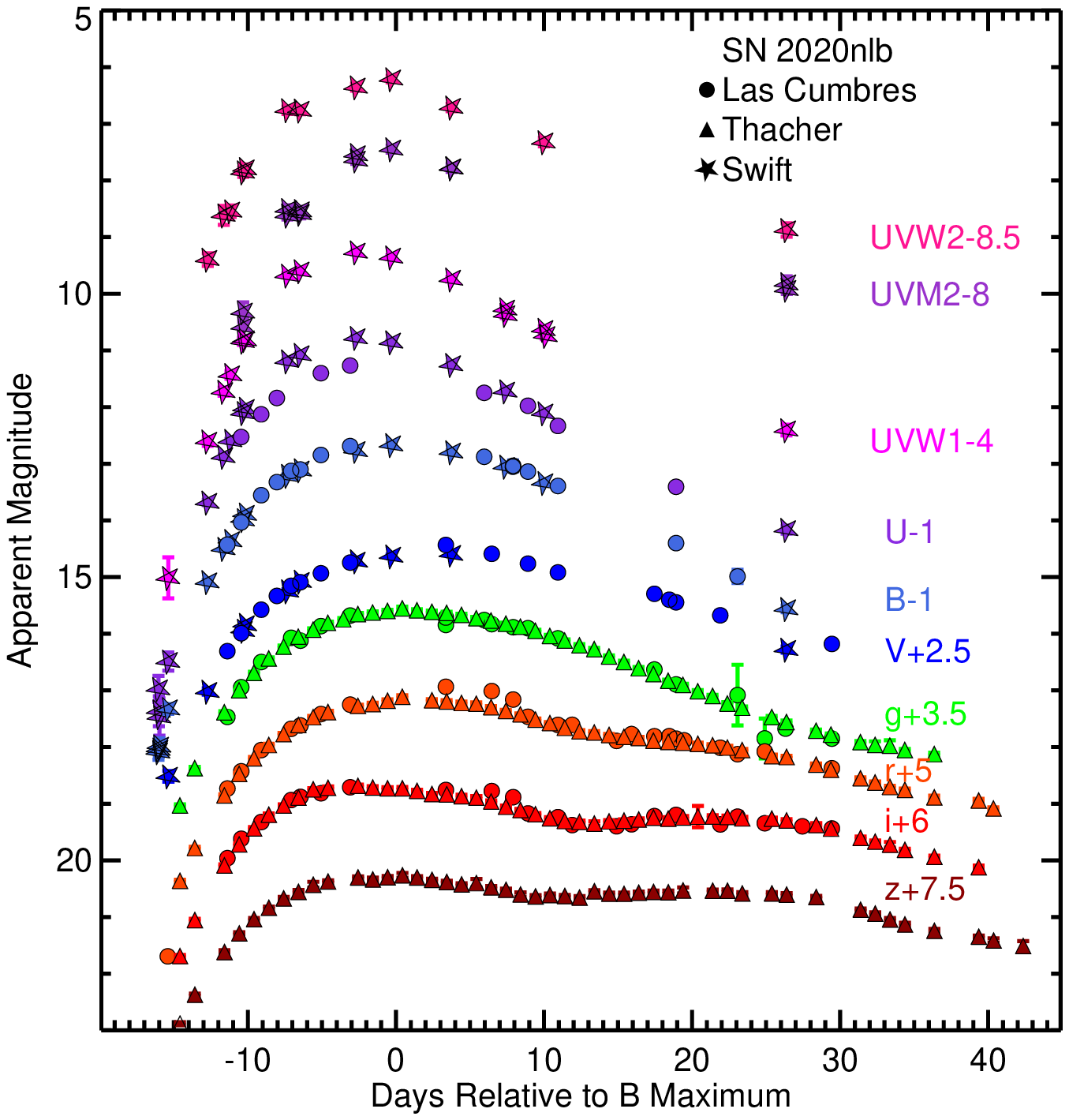}
\includegraphics[width=8.9cm]{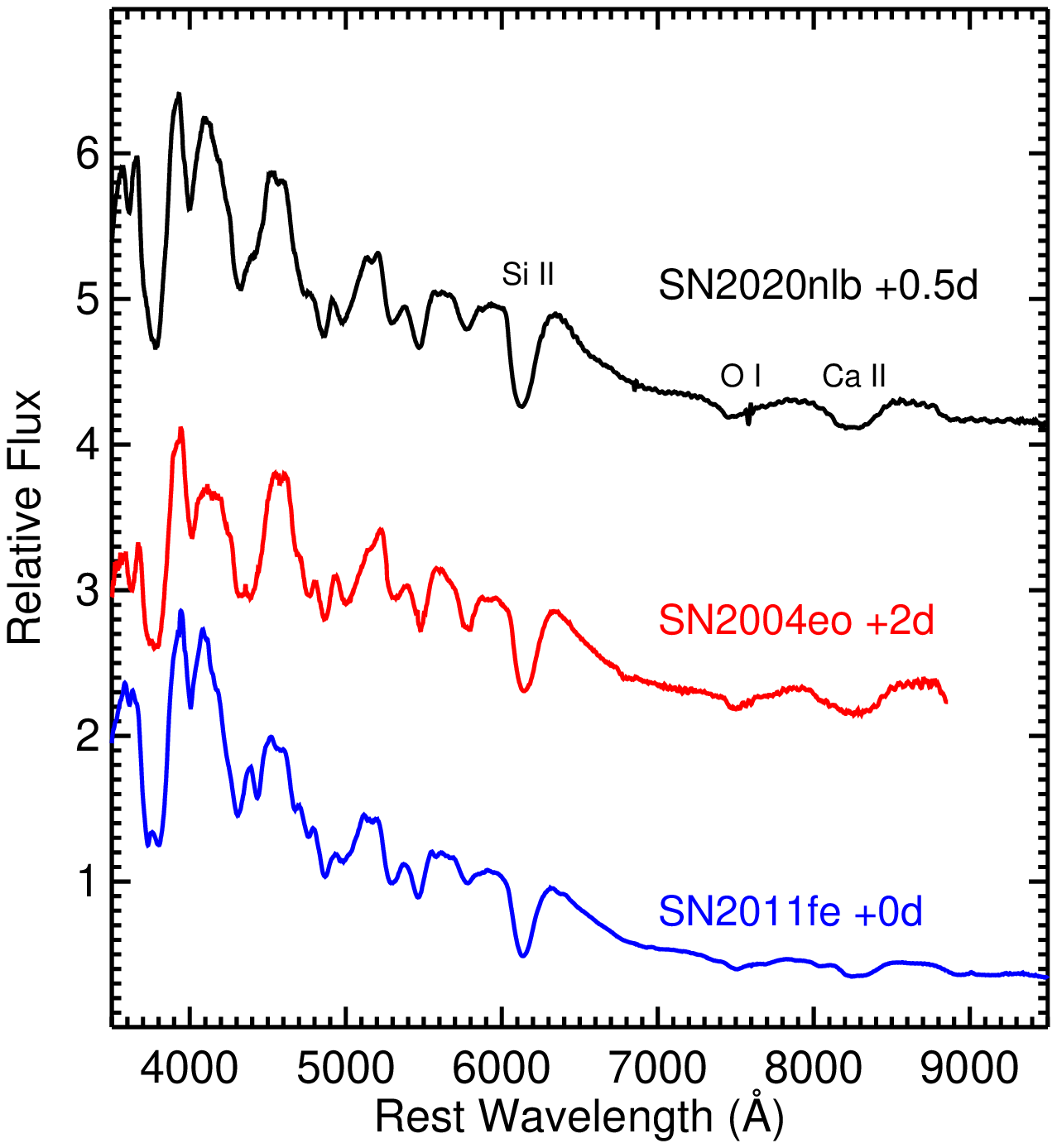}
\caption{Left: {\it Swift} UVOT, Thacher and Las Cumbres Observatory optical photometry of SN~2020nlb, spanning out to $\sim$30 days after $B$-band maximum.  The light curves have not been corrected for Milky Way extinction in this plot. The light curve will be made available in electronic format upon publication.  Right: The +0.5 day spectrum of SN~2020nlb, along with spectra of SN~2004eo \citep[+2 d;][]{Pastorello07} and the canonical `normal' type Ia SN~2011fe \citep[+0 d;][]{Pereira13}. \label{fig:lc_spec}}
\end{figure*}

In the right panel of Figure~\ref{fig:lc_spec}, we show a spectrum taken with the FLOYDS robotic spectrograph \citep{Brown13} at Faulkes Telescope North on 2020-07-12 06:32 UTC (+0.5 d with respect to $B$-band maximum), reduced with the pipeline described in \citet{Valenti14}.  Using the Supernova IDentification software package \citep[\texttt{SNID};][]{snid} we find that all reasonable matches correspond to SNe Ia near maximum light.  A particularly good match to SN~2004eo at +2 d with respect to $B$-band maximum was found, with similar \ion{Si}{2} and \ion{O}{1} strengths between the two events. SN~2004eo also has a similar, relatively fast light curve decline rate ($\Delta$m$_{15}$($B$)=1.45 mag; \citealt{Pastorello07}) as SN~2020nlb ($\Delta$m$_{15}$($B$)=1.29 mag).  We also compare the SN~2020nlb spectrum to the canonical `normal' SN Ia 2011fe at maximum light in Figure~\ref{fig:lc_spec} for illustrative purposes.

We measure a \ion{Si}{2} $\lambda$6355 velocity of 10,400$\pm$100 km s$^{-1}$ near maximum light, as well as pseudo-equivalent width (pEW) values of 118 \AA~and 24 \AA~for the \ion{Si}{2} $\lambda$6355 and $\lambda$5972 features, respectively, from the +0.5d FLOYDS spectrum.  In the standard Branch classification scheme \citep{Branch06}, SN~2020nlb is a Core Normal supernova, and belongs to the Normal Velocity class of SNe Ia as described in \citet{Wang09}.  

An additional low resolution optical spectrum was obtained with the Blue Channel spectrograph \citep{bluechan} at the MMT on 2021-01-28 21:06 UTC (+179 d), using the 300 l/mm grating and an exposure time of 3$\times$900 s.  We display this spectrum in Figure~\ref{fig:nebspec}, and use it in Section~\ref{sec:nebspec} to constrain any hydrogen or helium in the nebular phase.  To account for slit losses, we scale this spectrum to an $r$-band magnitude of 17.92, based on an interpolation of the Zwicky Transient Facility \citep[ZTF;][]{ztf} light curve at late times \citep[similar to][]{Sand18_2017cbv}.  


\section{X-Ray Observations And Analysis} \label{sec:obs}


\subsection{Swift-XRT}\label{subsec:swift}

The {\it Swift} \citep{Swift} X-Ray Telescope \citep[XRT;][]{xrt} observed both SN~2017cbv and SN~2020nlb extensively and we report flux limits here.  All XRT data was analyzed using HEASoft (v6.28) and corresponding calibration files.  Standard filtering and screening criteria were applied.  

We gathered {\it Swift} XRT data of SN~2017cbv taken between 2017 March 10.5 and April 15.46, corresponding to $\approx$1.5d and 37.5d from our adopted explosion epoch, with a total accumulated exposure time of 62.2 ks.  SN~2017cbv sits in a region of low background, and we obtain an unabsorbed (accounting for Galactic absorption) 3-$\sigma$ flux limit of $F$$<$7.7$\times$10$^{-15}$ ergs cm$^{-2}$s$^{-1}$ in the 0.3--10 keV energy range, corresponding to a 3-$\sigma$ luminosity limit of $L$ $<$ 1.6$\times$10$^{38}$ ergs s$^{-1}$ at a distance of 13.1 Mpc.

A sequence of {\it Swift} XRT data was taken of SN~2020nlb (alongside the UVOT data described in Section~\ref{sec:2020nlb}), starting on 2020 June 25.76 UTC.  We gathered all data taken through 2020 August 07.2 UTC ($\sim$43 days after our adopted explosion epoch), a total of 27.5 ks.  Unresolved, diffuse X-ray emission from the host elliptical galaxy M85 is apparent in the combined XRT data, which largely resolves into point sources in the {\it Chandra} data presented below.  At the position of SN~2020nlb, we find an unabsorbed 3-$\sigma$ flux limit of $F$$<$1.4$\times$10$^{-14}$ ergs cm$^{-2}$s$^{-1}$ in the 0.3--10 keV energy range, corresponding to a 3-$\sigma$ luminosity limit of $L$ $<$ 5.5$\times$10$^{38}$ ergs s$^{-1}$ at a distance of 17.9 Mpc.  

The XRT limits we obtain are a factor of $\sim$3--10 less stringent than the {\it Chandra} data we present in the next section, and for this reason we do not consider this data further as we constrain the circumbinary environment of SN~2017cbv and SN~2020nlb.

\begin{figure*}
\centering
\includegraphics[width=8.8cm]{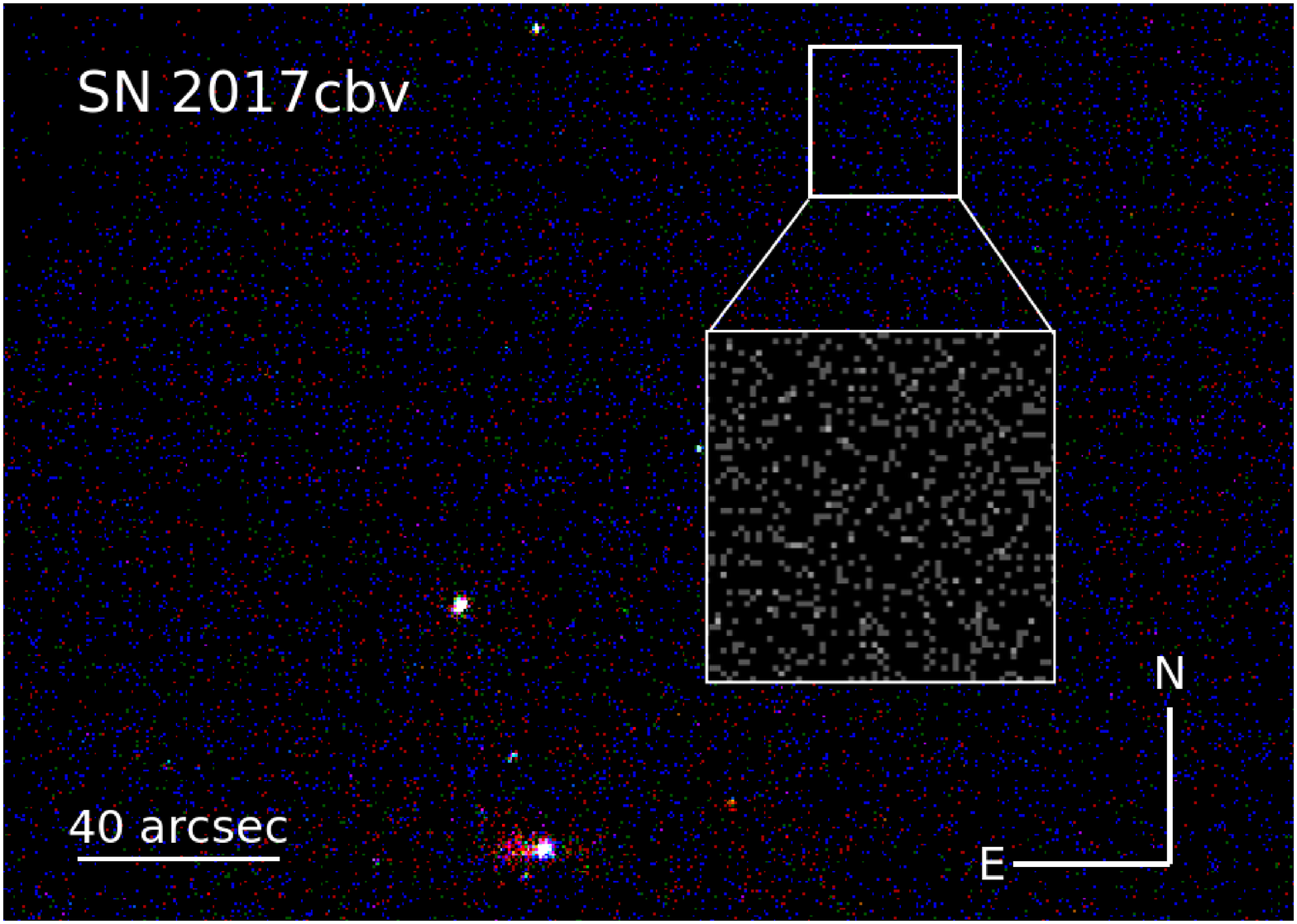}
\includegraphics[width=8.8cm]{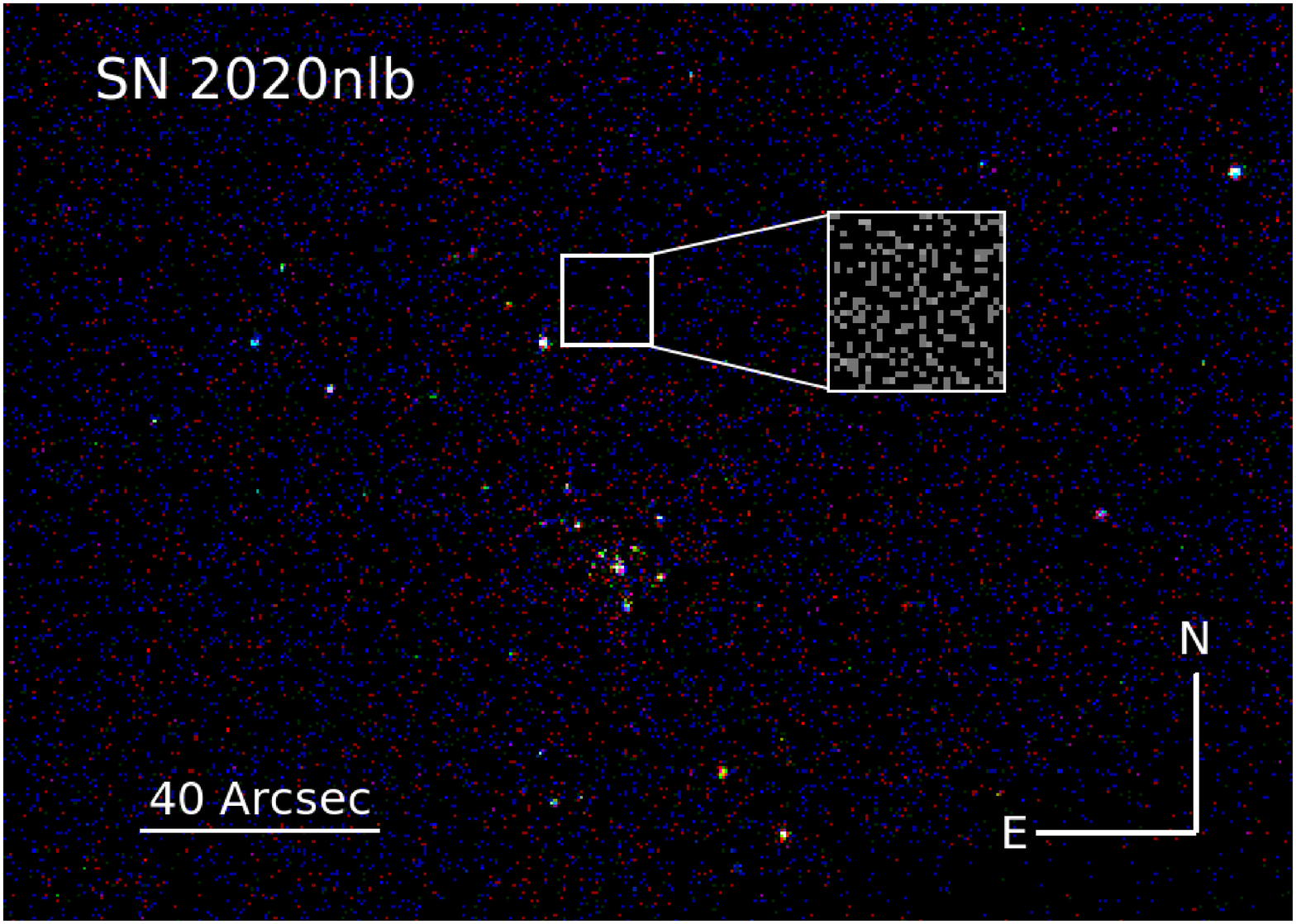}
\caption{False color {\it Chandra} X-ray images of both SN~2017cbv (left; $\delta$t=17.89 days after explosion) and SN~2020nlb (right; $\delta$t=15.79 days after explosion).  Red, green, and blue colors refer to soft (0.3--1.4 keV), medium (1.4--3 keV) and hard (3--10 keV) photons, respectively. The greyscale zoom-in on each panel is the full 0.3-10 keV {\it Chandra} energy range.  \label{fig:Xrayfig}}
\end{figure*}

\subsection{Chandra}\label{subsec:chandra}

Deep X-ray follow-up of both SN~2017cbv and SN~2020nlb were obtained with the {\it Chandra X-ray Observatory} under Director's Discretionary Time proposals.   All data were reduced within a conda \texttt{CIAO} (v4.12) environment using relevant calibration files (CALDB v4.9.1) and standard ACIS data filtering.  All X-ray count limits and confidence bounds are calculated assuming Poisson statistics, as described in \citet{Primini14}.   We present false color X-ray images, and zoom-ins on each SN location, in Figure~\ref{fig:Xrayfig}.  

Observations of SN~2017cbv began on March 27, 2017 (PI: Drout; Proposal 1850876; Obs ID 20055), and the total exposure time was 51~ks.  The midtime of the observations (MJD = 57839.789) corresponds to $\delta$t = 17.89 days with respect to the explosion epoch, or $-$1.1 days with respect to $B$-band maximum.  No X-ray source is detected with a 3--$\sigma$ upper limit of 1.8$\times$10$^{-4}$ counts s$^{-1}$ in the 0.3--10 keV energy band, which corresponds to a flux limit of $F$$<$2.2$\times$10$^{-15}$ ergs cm$^{-2}$ s$^{-1}$, assuming a power law model with spectral photon index $\Gamma$=2.  The unabsorbed flux limit is then $F$$<$2.6$\times$10$^{-15}$ ergs cm$^{-2}$ s$^{-1}$ (N$_H$ = 8.014$\times$10$^{20}$ cm$^{-2}$; 0.3--10 keV energy band), corresponding to a 3-$\sigma$ luminosity limit of $L$ $<$ 5.4$\times$10$^{37}$ ergs s$^{-1}$ at a distance of 13.1 Mpc.

{\it Chandra} observations of SN~2020nlb (PI: Sand; Proposal 21508740; Obs ID 23314, 23315)\footnote{\url{https://doi.org/10.25574/23314},~\url{https://doi.org/10.25574/23315}, respectively} were split into two blocks for a total exposure time of 75 ks.  The first observation (totaling 63 ks) began on 2020 July 09 23:58 (UT), while the second observation (12 ks) began $\sim$1.5 days after the end of the first on 2020 July 12 04:15. We take the weighted average time of these two exposures as the effective epoch of the observations, MJD 59040.7, which corresponds to $\delta t$=15.79 days with respect to the assumed explosion epoch, or $-$1.10 days with respect to $B$-band maximum.  For our analysis, we extract the individual source spectra and use the \texttt{CIAO}/{\it specextract} task for combination of the spectra and appropriate response files.  We have also combined the two observations into a single event map using the \texttt{CIAO}/{\it merge\_obs} task for illustrative purposes in Figure~\ref{fig:Xrayfig}, to show that no source is present.  For the combined spectrum, at the position of SN~2020nlb, we find a 3-$\sigma$ flux limit of $F$$<$1.3$\times$10$^{-15}$ ergs cm$^{-2}$ s$^{-1}$ in the 0.3--10 keV energy range, assuming a power law model with spectral photon index $\Gamma$=2.  The unabsorbed flux limit is then $F$$<$1.4$\times$10$^{-15}$ ergs cm$^{-2}$ s$^{-1}$ (N$_H$ = 2.49$\times$10$^{20}$ cm$^{-2}$; 0.3--10 keV energy band), and the 3-$\sigma$ luminosity limit is $L$ $<$ 5.4$\times$10$^{37}$ ergs s$^{-1}$ at a distance of 17.9 Mpc.  

\begin{figure}
\centering
\includegraphics[width=\columnwidth]{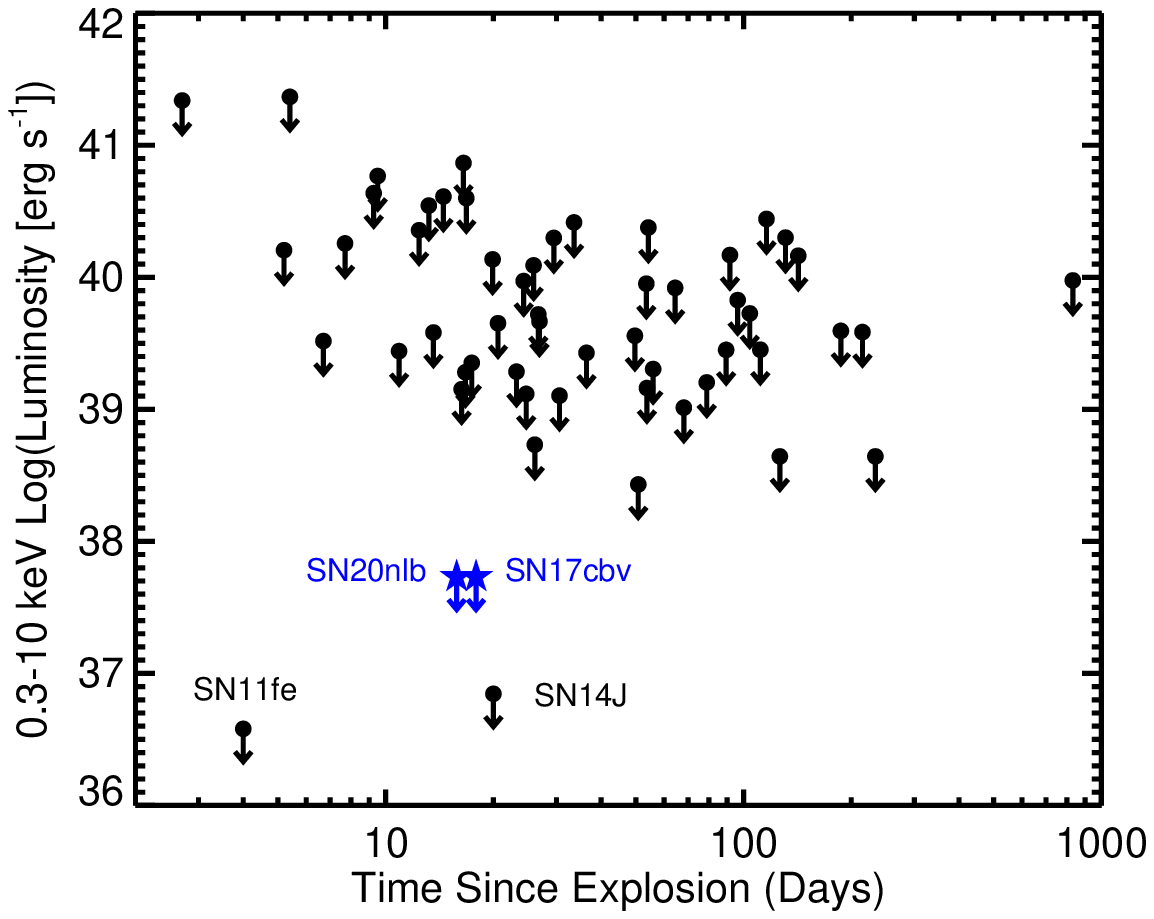}
\caption{X-ray luminosity limits of normal SNe Ia from the compilation of \citet{Russell12}, along with deep {\it Chandra} observations of SN 2011fe \citep{Margutti12} and SN~2014J \citep{Margutti14}.  The new {\it Chandra} X-ray limits for SN~2020nlb and SN~2017cbv are also shown as blue stars, and are  the deepest data taken of SNe Ia beyond D=10 Mpc.  Note that the  data from \citet{Russell12} is generally derived from the combination of {\it Swift} XRT data taken across many epochs over $\sim$30 days, which is not well-represented by the points in the plot.    \label{fig:Xraylum}}
\end{figure}

We place the {\it Chandra} X-ray luminosity limits of SN~2017cbv and SN~2020nlb in context in Figure~\ref{fig:Xraylum}, in comparison to the majority of X-ray limits in the literature.  The data on SN~2017cbv and SN~2020nlb are amongst the most constraining obtained for any SNe Ia, just behind the very nearby SNe 2011fe and 2014J.  We use these limits in Section~\ref{sec:x_constrain} to constrain any CSM associated with these SNe.

\section{The Bolometric Luminosity}\label{sec:lum}

X-ray emission in low density, hydrogen stripped progenitor systems in the first $\sim$1--2 months after explosion is dominated by Inverse Compton scattering of photospheric photons by relativistic electrons accelerated by the SN shock \citep{Chevalier06,Margutti12}.  The Inverse Compton X-ray luminosity is proportional to the bolometric luminosity of the SN, 
and so bolometric luminosity  is a key ingredient of our analysis, which we discuss here.  

We adopt the pseudo-bolometric light curve of SN~2017cbv published in \citet[][see their Table 7]{Wang20},  who combined UV+optical+NIR photometry to construct a pseudo-bolometric light curve using the SNooPy light curve package \citep{Burns11,Burns14}; we refer the reader to that work for details.  Interpolating the \citet{Wang20} bolometric light curve to the {\it Chandra} X-ray epoch yields $L_{bol}$=1.46$\times$10$^{43}$ ergs s$^{-1}$ at our adopted distance of 13.1 Mpc.

Based on our UV+optical light curve of SN~2020nlb presented in Section~\ref{sec:SN20nlb_lc}, we construct a pseudo-bolometric light curve using the `direct' method in the SNooPy light curve package.  We use the $UVW1$ and $UVW2$ filters from {\it Swift} and the Las Cumbres $UBVgri$ light curves for our analysis (which encompasses a broad wavelength range on one telescope system), ignoring the $UVM2$ filter as its light curve was lower signal to noise and sparser, making interpolation between epochs difficult.  The `direct' method takes the flux in each filter and integrates over all filters, interpolating when necessary when a given filter is missing.  The program also takes into account the extinction (where we are only considering Milky Way extinction, see Section 2.2.1) and distance to the supernova. We extrapolate the flux in the near infrared by assuming a Rayleigh-Jeans tail, although this only has a small effect on our results.  We also experimented with the `SED' method in SNooPy, using the \citet{Hsiao07} spectral energy distribution templates, and obtained results consistent with the direct method to within $\sim$4\%.  Given this, we utilize the `direct' results and find $L_{bol}$=1.0$\times$10$^{43}$ ergs s$^{-1}$ at the epoch of the {\it Chandra} X-ray observations (MJD 59040.7).

We plot both of the bolometric light curves in the bottom panels of Figure~\ref{fig:lums}, and highlight the luminosity at the epoch of the {\it Chandra} observations.

\begin{figure*}
\begin{center}
\includegraphics[width=15cm]{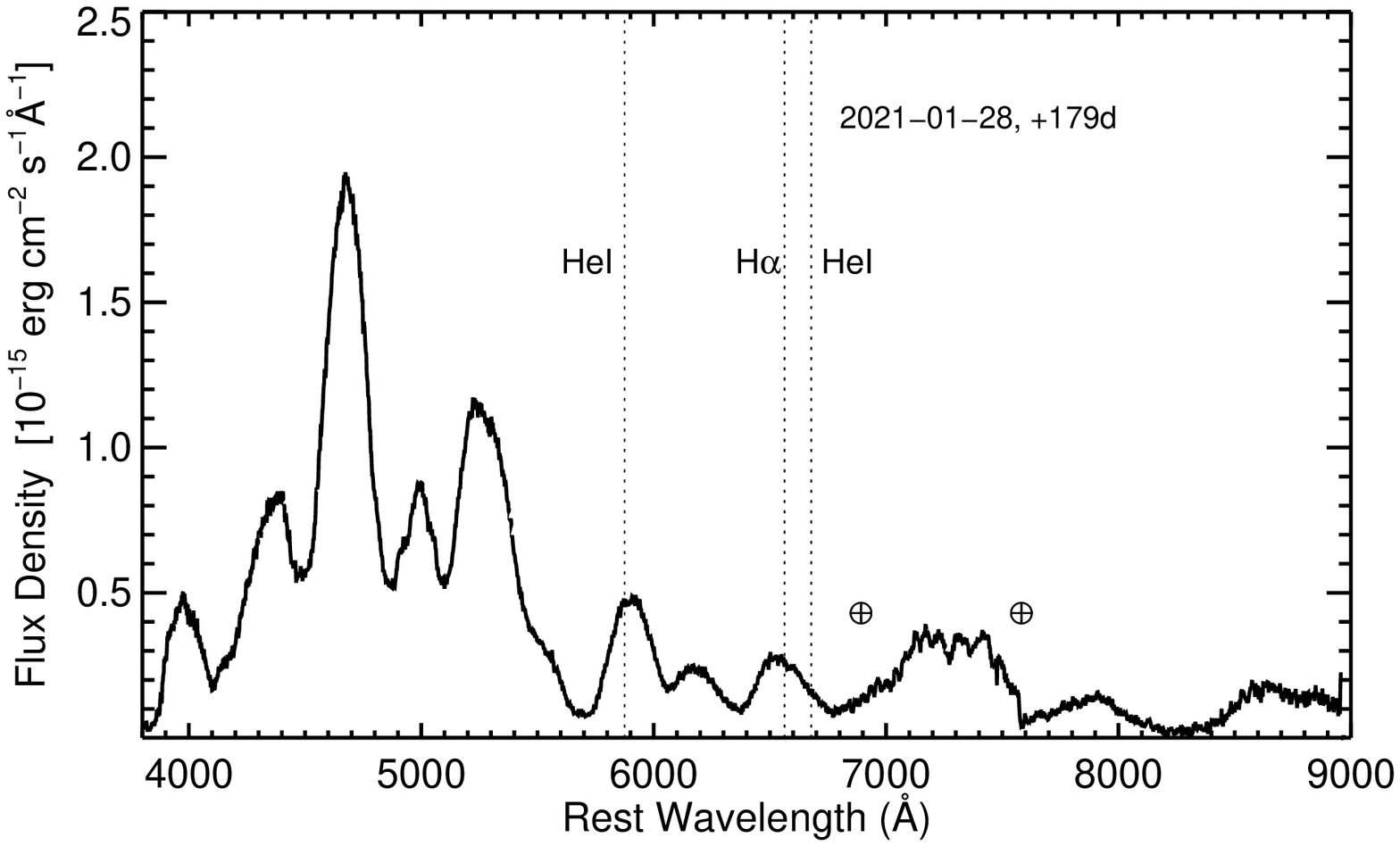}
\includegraphics[scale=0.5]{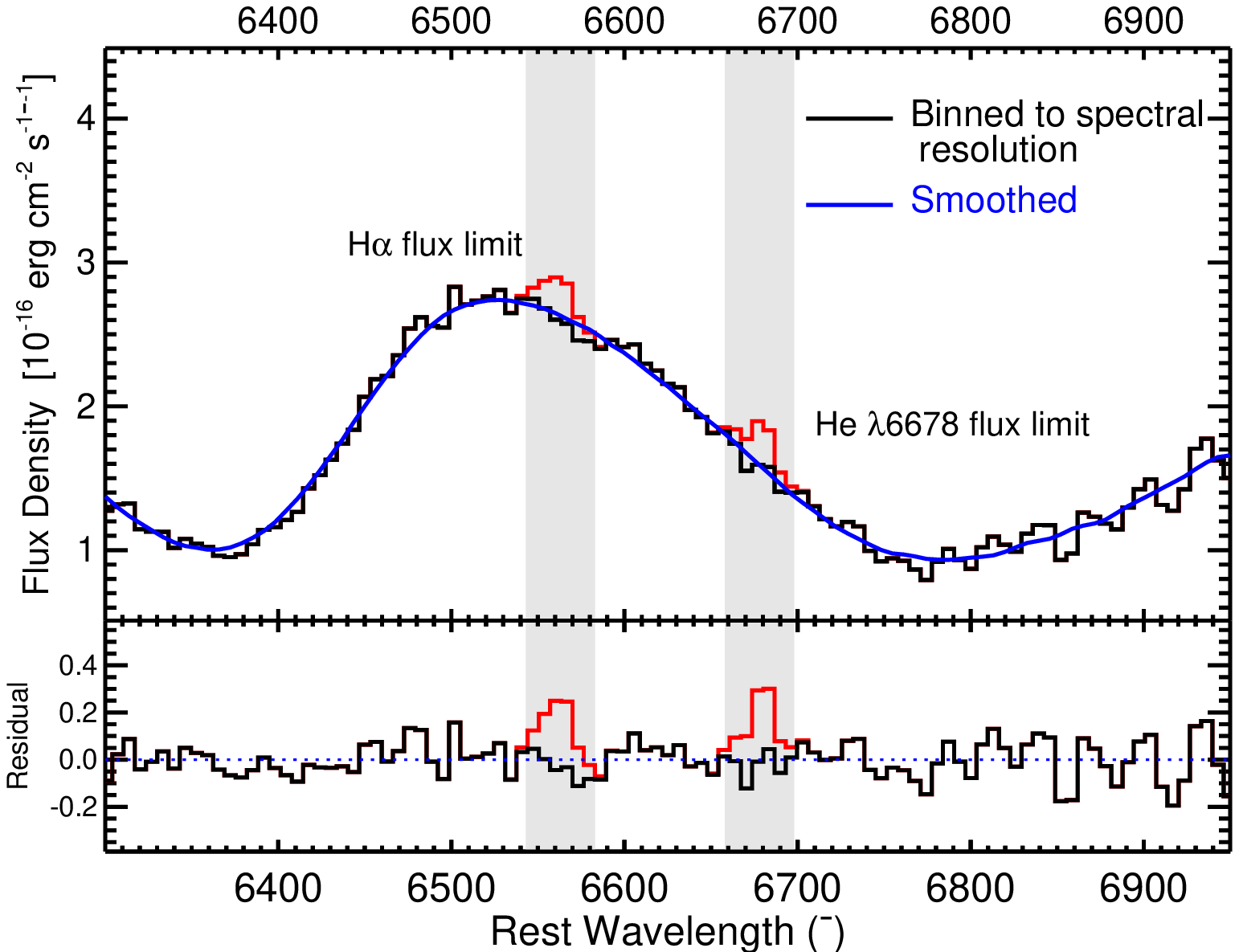}
\includegraphics[scale=0.5]{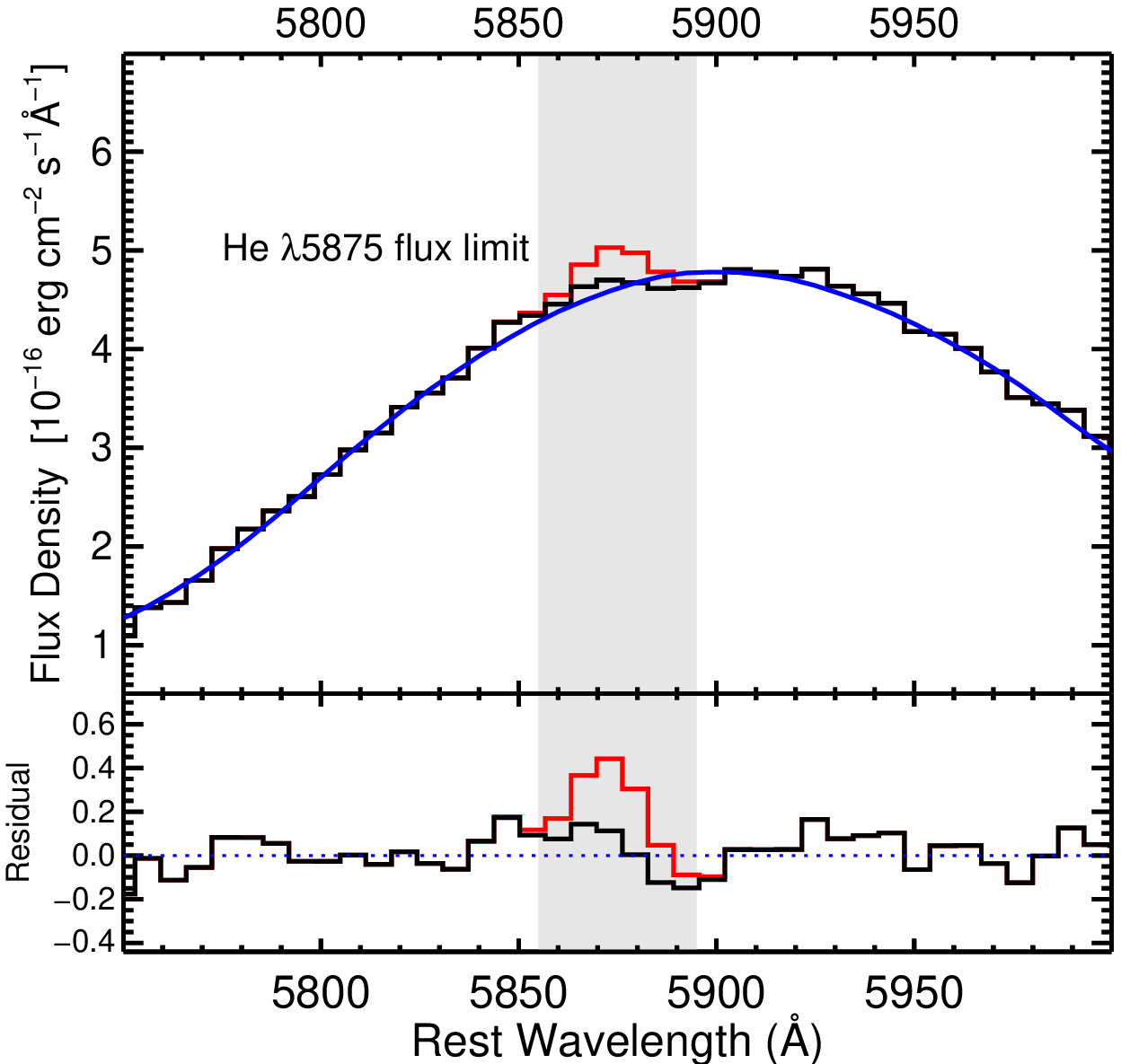}
\caption{Top -- Nebular-phase spectrum of SN~2020nlb, flux calibrated and corrected for MW extinction.  We mark the locations of the A- and B- band telluric features.  Bottom -- Zoom in on the region around H$\alpha$ and \ion{He}{1} $\lambda$6678 (left) as well as \ion{He}{1} $\lambda$5875 (right), where we constrain any narrow emission feature by implanting simulated lines.  The black histogram is the data binned to the resolution of the spectrum, while the blue shows the data smoothed by a second order Savitsky-Golay filter.  The bottom panel on each plot shows the difference between the original spectrum and the smoothed version.  The red histogram is an implanted emission line feature with a peak flux four times the rms, representing the detection limit for an hydrogen or helium emission line.  The gray shaded regions correspond to 1000 km s$^{-1}$ around the rest wavelength of the hydrogen and helium lines. \label{fig:nebspec}}
\end{center}
\end{figure*}

\begin{figure*}
\centering
\includegraphics[width=8.8cm]{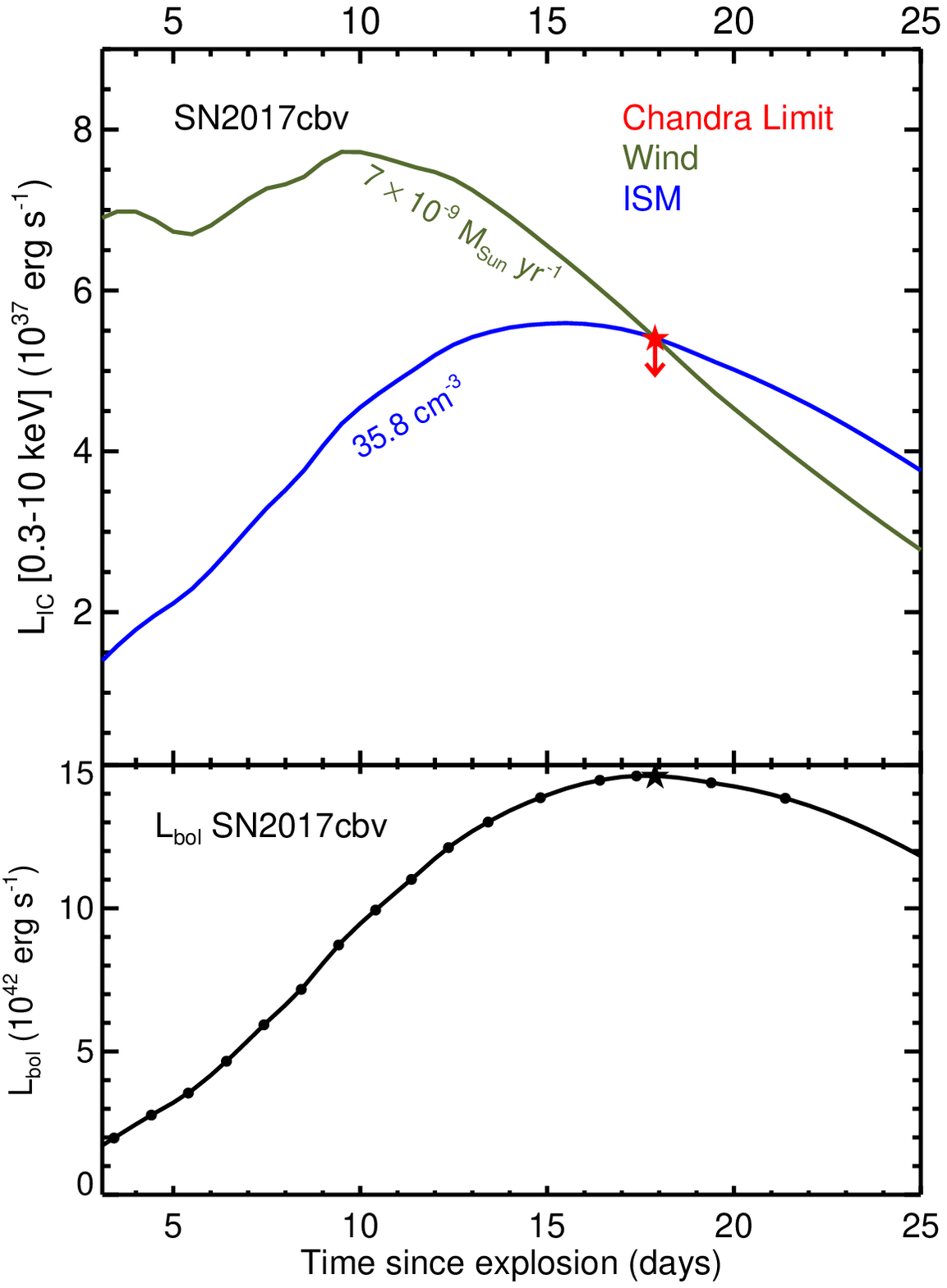}
\includegraphics[width=8.8cm]{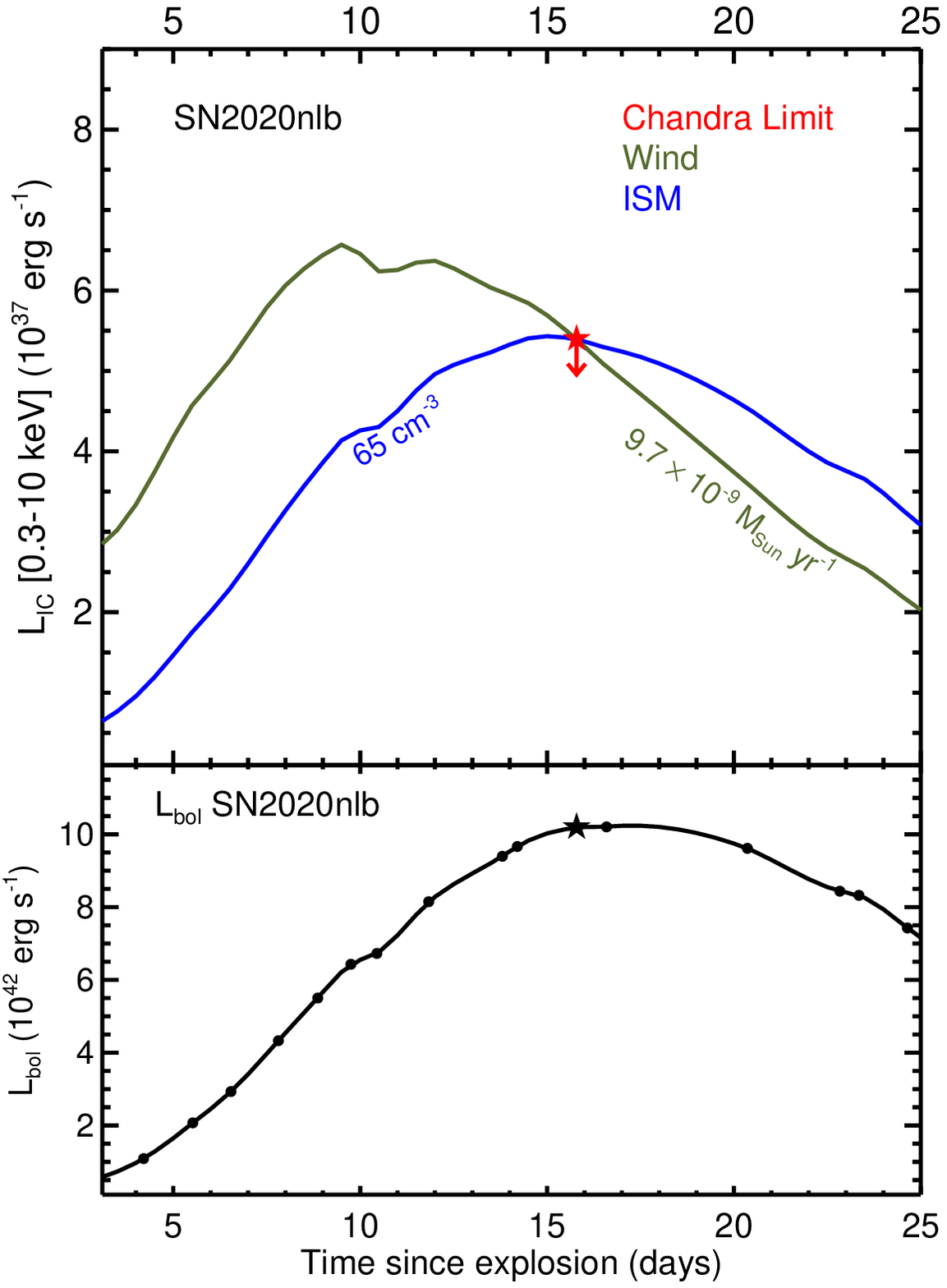}
\caption{Top panels -- Expected maximum inverse Compton X-ray luminosity  in the case of wind (green, with an assumed $v_w$=100 km s$^{-1}$) and ISM-like (blue) environments for SN~2017cbv (left) and SN~2020nlb (right), based on the {\it Chandra} X-ray limits (red upper limits) for each.  Bottom panels -- Bolometric light curves for SN 2017cbv (data from \citealt{Wang20}) and SN~2020nlb (this work).  The interpolated bolometric luminosity of the SN at the epoch of the {\it Chandra} observations is marked with a star symbol. \label{fig:lums}}
\end{figure*}

\begin{figure*}
\centering
\includegraphics[width=14cm]{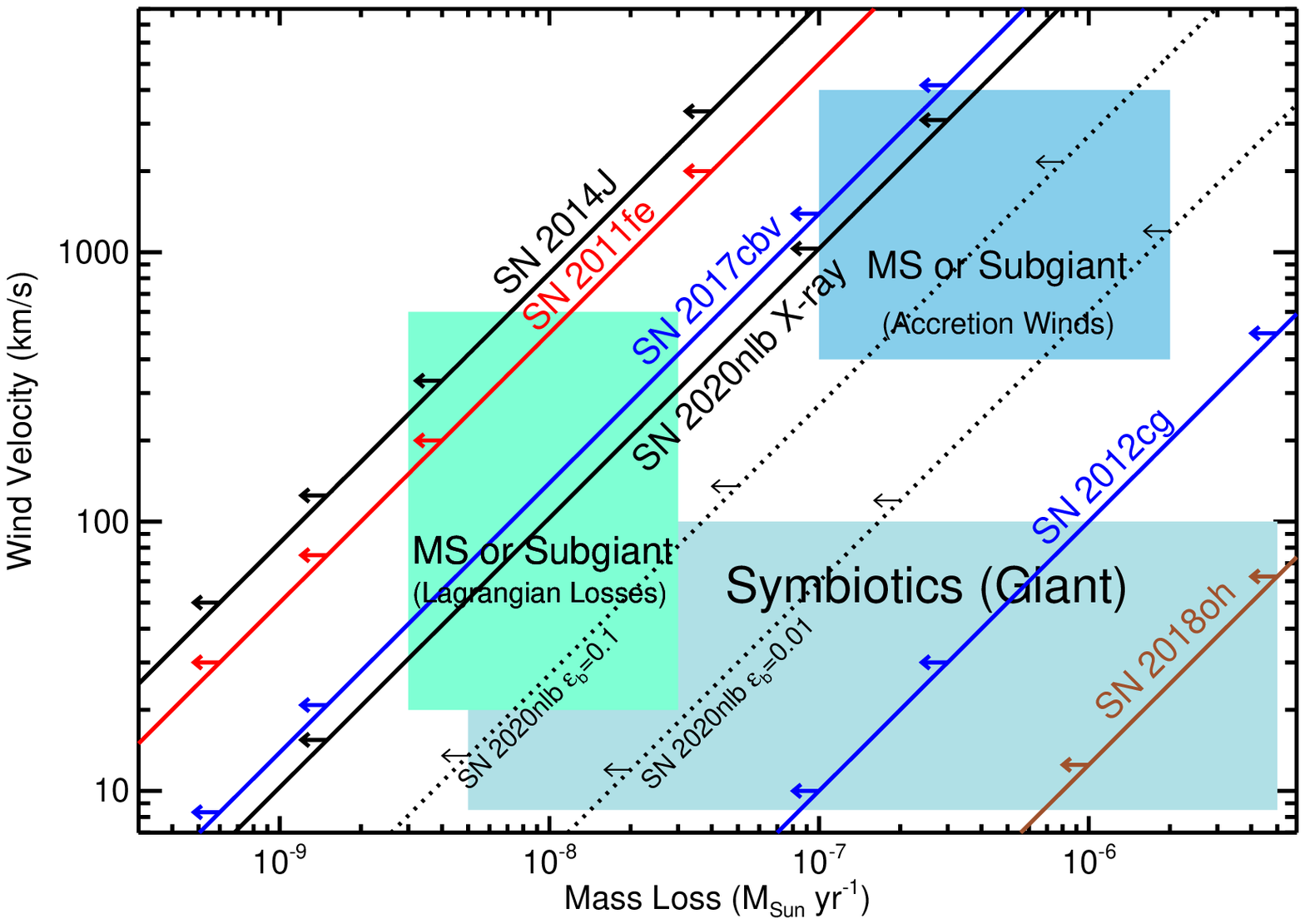}
\caption{X-ray constraints on a wind-like CSM for SN~2017cbv (blue line) and SN~2020nlb (solid black line) with respect to other prominent results in the literature.  Regions of parameter space not ruled out are towards the upper left of the plot, in the direction of the arrows.  The color blocked regions represent areas of parameter space where plausible SN Ia progenitors and their expected CSM would inhabit, which we discuss further in Section~\ref{sec:discussion}.  Limits obtained for other prominent SNe Ia include SN 2011fe \citep{Margutti12}, SN~2014J \citep{Margutti14}, SN~2012cg \citep{Shappee18} and SN~2018oh \citep{Shappee19}. The data from \citet{Russell12}, as seen in Figure~\ref{fig:Xraylum}, all lie to the right of SN~2012cg in this plot.  We also plot our radio CSM constraints for SN~2020nlb (dotted lines; Section~\ref{sec:radio}) assuming $\epsilon_b$=0.01 and 0.1 for the fraction of shock energy shared by the amplified magnetic field; these observations alone rule out portions of the parameter space occupied by symbiotics and accretion winds. \label{fig:CSM}}
\end{figure*}

\section{SN~2020nlb Progenitor Constraints from Nebular Spectroscopy}\label{sec:nebspec}

We measure complementary constraints on the progenitor system and environment of SN~2020nlb using the flux calibrated, extinction and redshift-corrected nebular spectrum (+179d with respect to $B$-band maximum, and +196d with respect to our adopted explosion epoch of MJD 59024.91) presented in Section~\ref{sec:sn20nlb_spec} and shown in Figure~\ref{fig:nebspec}, to go along with our primary X-ray results in the next section.  We remind the reader that we have scaled this spectrum to $r$=17.92 mag to match the late time ZTF photometry to account for nonphotometric conditions and slit losses.  Using similar techniques, strong constraints on H and He emission have already been placed on SN~2017cbv, using the models of \citet{Boty18}, with limits of $M_{\rm H} < 1 \times 10^{-4}$ $\rm M_{\odot}$ and $M_{\rm He} < 5 \times 10^{-4}$ $\rm M_{\odot}$, respectively; this is $\sim$3 orders of magnitude below expectations for the single degenerate scenario \citep{Sand18_2017cbv}.

If the progenitor system of SN~2020nlb had a nondegenerate companion star, then models predict that the SN ejecta will impact the companion, and manifest as narrow hydrogen (or helium) emission lines with FWHM$\approx$1000 km s$^{-1}$ at late times \citep[e.g.,][among others]{Marietta00,Mattila05,Pan10,Pan12,Liu12,Liu13,Lundqvist13,Boty18,Dessart20}. The models for the emission from stripped material expect $\gtrsim$0.1 $M_{\odot}$ of stripped hydrogen, but have considerable diversity in the strength and shape of the observed emission line, and depend on the details of the explosion and radiative transfer physics employed. Here we will rely on the latest radiative transfer modeling and predictions from \cite{Boty18} and \cite{Dessart20}, and refer the reader to those works for details.

Careful visual inspection of Figure~\ref{fig:nebspec} reveals no hydrogen or helium  emission features, including from the host galaxy.  To set quantitative limits on narrow H$\alpha$ as well as \ion{He}{1} $\lambda$5875 \AA\ and $\lambda$6678 \AA\ emission, we mimic the methodology of \citet{Sand18_2017cbv,Sand19}, which we briefly describe here.  We take the flux-calibrated, extinction and redshift-corrected spectrum and bin to the native resolution, $\approx$6.5 \AA.  We then establish a `continuum level' around the hydrogen and helium line wavelengths by smoothing the spectrum on scales larger than the expected emission (FWHM$\approx$1000 km s$^{-1}$) using a second-order Savitsky-Golay filter with a width of 190 \AA.  We experimented with various filter widths in order to best recover simulated emission line features in our data.  Any hydrogen or helium emission line feature of the width we are interested in would be apparent in the difference between the smoothed and un-smoothed spectrum, which we refer to as the residual spectrum.

To estimate the maximum H$\alpha$ (or helium) emission that could go undetected, we  directly implant emission lines into our data.  We assume a line width of FWHM$\approx$1000 km s$^{-1}$ and a peak flux that is four times the root mean square of the residual spectrum.  This results in an H$\alpha$ flux limit of 7.1$\times$10$^{-16}$ erg s$^{-1}$ cm$^{-2}$, and a luminosity limit of 2.7$\times$10$^{37}$ ergs s$^{-1}$ at a distance of D=17.9 Mpc.  Similarly, we find a 
\ion{He}{1} $\lambda$6678 \AA\ ($\lambda$5875 \AA) flux limit of 7.1$\times$10$^{-16}$ erg s$^{-1}$ cm$^{-2}$ (7.9$\times$10$^{-16}$ erg s$^{-1}$ cm$^{-2}$), leading to a luminosity limit of 2.7$\times$10$^{37}$ ergs s$^{-1}$ (3.0$\times$10$^{37}$ ergs s$^{-1}$).  \citet{Dessart20} suggest that limits on the equivalent width of the H$\alpha$ line may also be an effective way to obtain stripped-mass limits.  Given this, we have also followed the \citet{Dessart20} prescription for obtaining equivalent width limits (using their Equation B.1), and find EQW(H$\alpha$)$<$5.3 \AA.  We illustrate our detection limits in the bottom panels of Figure~\ref{fig:nebspec}.

The recent 3D radiation transport results of \citet{Boty18} presented simulated SN Ia spectra at 200 days after explosion (well-matched to our spectrum at +196 days after explosion, which we make no further adjustments to) derived from the SN Ia ejecta-companion interaction simulations of \citet{Boehner17}, which utilized a spherically symmetric W7 explosion model \citep{Nomoto84,Thielemann86}.  These simulated spectra show strong hydrogen and helium nebular emission (FWHM$\approx$1000 km s$^{-1}$, shifted by up to $\sim$10 \AA\ from rest) with $L_{H\alpha}$$\approx$4.5--15.7$\times$10$^{39}$ ergs s$^{-1}$ for their main sequence, subgiant and red giant companion star models, corresponding to $M_{\rm strip}$$\sim$0.2--0.4 $M_{\odot}$.  Given our H$\alpha$ luminosity limit of 2.7$\times$10$^{37}$ ergs s$^{-1}$, we rule out the basic predictions of \citet{Boty18} in SN~2020nlb by over two orders of magnitude.  To approximate the effects of smaller stripped hydrogen masses (due to weaker explosions, wider companion separations or inhomogeneities in the ejecta structure), \citet{Boty18} varied the hydrogen density in their fiducial main sequence companion model, fitting a quadratic to the relation between stripped hydrogen mass and the H$\alpha$ luminosity (see their Equation 1, but note the typographical error corrected in \citealt{Sand18_2017cbv}).  Using this formula, we derive a stripped hydrogen mass limit of $M_{\rm strip}$=7$\times$10$^{-4}$ M$_{\odot}$.

Further radiative transfer calculations of SN Ia ejecta enclosing stripped material from a nondegenerate companion star were recently performed by \citet{Dessart20}. These 1D calculations  include several different delayed detonation models \citep[DDC0, DDC15, and DDC25;][]{Blondin13} and a sub-Chandrasekhar model \citep[SCH3p5;][]{Blondin17}, expanding beyond the focus on W7 explosions in previous work.  They also include non-LTE physics and optical depth effects, and cover stripped hydrogen masses from $M_{\rm strip}$$\sim$0.001--0.5 M$_{\odot}$ (although we note that masses in the $M_{\rm strip}$$\sim$0.1--0.5 M$_{\odot}$ range are what is seen in multi-dimensional hydrodynamic simulations, as mentioned above).  We estimate hydrogen mass limits of $M_{\rm strip}$=1--2$\times$10$^{-3}$ M$_{\odot}$ based on these models, where the range includes the three explosion models with tabulated results in the Appendix of \citet{Dessart20}.  When we use our equivalent width limit for H$\alpha$ (EQW(H$\alpha$)$<$5.3 \AA) instead of our luminosity limit, we obtain nearly identical hydrogen mass limits of $M_{\rm strip}$=1--3$\times$10$^{-3}$ M$_{\odot}$.  All of these mass limits are $\sim$2 orders of magnitude lower than expectations from single degenerate models.

Helium stars are also plausible white dwarf companions \citep[e.g.][]{Iben84}, and models predict that such progenitor systems should yield $M_{\rm strip}$$\approx$0.002--0.06 M$_{\odot}$ of stripped helium-rich material which ultimately may be visible in nebular spectra \citep[e.g.][]{Pan12,Liu13}.  Unfortunately, few radiative transport simulations of the helium star scenario have been done, and so it is difficult to translate helium luminosity limits to stripped helium masses.  As an approximate solution, we utilize the \citet{Boty18} model where they replaced hydrogen with helium in their simulations, and assume that the helium mass falls off with luminosity with the same quadratic form as for the stripped hydrogen models (see also \citealt{Sand18_2017cbv}).  From this we infer a limit of $M_{\rm strip}$$\sim$4$\times$10$^{-3}$ M$_{\odot}$, although this value should be treated with caution.  We note that the recent radiative transfer work of \citet{Dessart20} produce weak or ambiguous helium lines in their simulated spectra, aside from the stronger \ion{He}{1} $\lambda$10830 \AA~line.  Thus, our helium limits correspond to the lower end of expectations for helium companion stars in the single degenerate model, but further work is needed to produce robust helium line predictions.

In summary, our limits on stripped hydrogen from the single degenerate scenario are $\sim$2--3 orders of magnitude below expectations, setting a strong constraint on this scenario, modulo current model limitations.  We further rule out most helium companion star scenarios, although our helium limit of $M_{\rm strip}$$\sim$4$\times$10$^{-3}$ M$_{\odot}$ does overlap with some predictions, so we cannot completely rule out this channel.  Our hydrogen limits are similar to or stronger than the H$\alpha$ recently seen in the fast declining SNe Ia ASASSN-18tb ($L_{H\alpha}$=2.2$\times$10$^{38}$ ergs s$^{-1}$; \citealt{Kollmeier19}), SN~2018cqj ($L_{H\alpha}$=3.8$\times$10$^{37}$ ergs s$^{-1}$; \citealt{Prieto20}), and SN~2016jae ($L_{H\alpha}$=1.6--3.0$\times$10$^{38}$ ergs s$^{-1}$; \citealt{Elias21}).  As an aside, our hydrogen limits also rule out any hydrogen-rich CSM (although the model expectations are less clear here) which may be interacting with the ejecta at this epoch.  Outside of the single degenerate scenario, such CSM may plausibly be associated with a giant planet \citep{Soker19} or a nondegenerate tertiary star \citep{Thompson11,Kushnir13,Vallely19}, which does not seem to be the case for SN~2020nlb.

\section{SN~2020nlb Constraints on Circumstellar Interaction from Radio Observations} \label{sec:radio}

Radio emission can independently probe the presence of circumstellar material (CSM) because interaction of the SN ejecta with this CSM accelerates electrons to relativistic energies and amplifies the ambient magnetic field, producing radio synchrotron emission \citep{Chevalier1982, Chevalier1984, Chevalier1998}. Simple models of radio emission have provided constraints on the CSM environment and progenitor properties for both core-collapse \citep[e.g.][]{Ryder2004, Soderberg2006, Chevalier2006, Weiler2007, Salas2013, Bostroem19} and Type Ia SNe \citep{Panagia2006, Chomiuk2016}. Radio emission is yet to be detected from a Type Ia SN, but non-detections have provided stringent constraints on their progenitor scenarios \citep{Chomiuk2016}, particularly for nearby events \citep{Horesh2012, Chomiuk2012, Torres2014, Pellegrino2020, Lundqvist2020, Burke21}. 

Below we describe our VLA observations of SN 2020nlb. SN 2017cbv is outside the declination limit of the VLA, and we are not aware of any radio observations  that were associated with this SN.

\subsection{Observation}
A radio observation of SN 2020nlb was obtained with the Karl G.\ Jansky Very Large Array (VLA) on 2020 July 7 at 21:49, which is 14.6 days since explosion (derived in Section \ref{sec:2020nlb}). The observation block was 1-hr long, with 37.5 mins on-source time for SN 2020nlb. Observations were taken in C-band (4--8 GHz) in the B-configuration of the VLA (DDT: 20A-577, PI: S. Sarbadhicary).\ The observations were obtained in wide-band continuum mode, yielding 4 GHz of bandwidth sampled by 32 spectral windows, each 128 MHz wide sampled by 1 MHz-wide channels with two polarizations. We used 3C286 as our flux, delay and bandpass calibrator, and J1224+2122 as our complex gain calibrator. 
Data were calibrated with the VLA CASA calibration pipeline (version 5.6.1-8), which iteratively flags corrupted measurements, applies corrections from the online system (e.g. antenna positions), and applies delay, flux density, bandpass and complex gain calibrations. We then imaged the calibrated visibility dataset with \texttt{tclean} in CASA. We used multi-term, multi-frequency synthesis as our deconvolution algorithm (set with \texttt{deconvolver=`mtmfs'} in \texttt{tclean}), which approximates the full 4-8 GHz wide-band spectral structure of the sky brightness distribution as a Taylor-series expansion about a reference frequency (in our case, 6 GHz) in order to reduce frequency-dependent artifacts during deconvolution \citep{Rau2011}. We set \texttt{nterms=2} which uses the first two Taylor terms to create images of intensity (Stokes-I) and spectral index. We sampled the synthesized beam with 3-4 pixels, and imaged out to 11.2\arcmin\ (about 12$\%$ sensitivity level of primary beam) to deconvolve any outlying bright sources and mitigate their sidelobes at the primary beam center. Gridding was carried out with the W-projection algorithm (\texttt{gridder=wproject}) with 16 w-planes. Images were weighted with the Briggs weighting scheme (\texttt{weighting=briggs}) using a robust value of 0 to balance point-source sensitivity with high angular resolution and low sidelobe contamination between sources. The final image has a spatial resolution of 0.9\arcsec $\times$ 0.8\arcsec (or roughly $79\times70$ pc), and an RMS of about 5 $\mu$Jy/bm, which is within 25$\%$ of the expected thermal noise level of our observation.\footnote{\url{https://obs.vla.nrao.edu/ect/}}

No radio source was detected at the site of SN 2020nlb in the cleaned, deconvolved 6-GHz image at the 3$\sigma$ level. The flux at the SN location is $6.4~\mu$Jy/beam, and the RMS noise in a circular region 7\arcsec\ across around the SN location is $4.5~\mu$Jy/beam. We therefore assume a flux density upper limit of 19.7 $\mu$Jy/beam, which is equal to the flux density at the SN location plus three times the RMS noise. At a distance of 17.9 Mpc, this corresponds to a 6 GHz luminosity of $7.6 \times 10^{24}$
ergs s$^{-1}$ Hz$^{-1}$.

\subsection{Analysis}
The luminosity upper limit can shed some light on the CSM around SN 2020nlb similar to the methodology in \cite{Chomiuk2012} and \cite{Chomiuk2016}. We assume SN 2020nlb was surrounded by the \cite{Chevalier1982} model of a CSM, produced by steady mass loss from the progenitor, i.e. $\rho_{csm} = \dot{M}/4 \pi r^2 v_w$ (where $\rho_{csm}$ is the CSM density in g cm$^{-3}$, $\dot{M}$ is the mass-loss rate from the progenitor, $r$ is the distance from the progenitor and $v_w$ is the wind velocity). Assuming a standard SN Ia explosion with 10$^{51}$ ergs kinetic energy and 1.4 M$_{\odot}$ ejecta mass, we obtain a mass-loss rate upper limit of $\dot{M} < (3.7-16.7) \times 10^{-9}$ M$_{\odot}$ yr$^{-1}$, assuming $v_w$=10 km s$^{-1}$. The range of mass-loss rates reflect the uncertainty in the parameter $\epsilon_B$, the fraction of shock energy shared by the amplified magnetic field, with typical values in the range 0.01--0.1 for SNe \citep{Chomiuk2012}.
These limits are compared with the mass-loss rate parameter space of single-degenerate models as defined in \cite{Chomiuk2012} in Figure \ref{fig:CSM}. We find that our limits are deep enough to rule out red-giant companions (symbiotic systems), characterized by slow winds of 10-100 km s$^{-1}$ and mass-loss rates of 10$^{-6}$--10$^{-8}$ M$_{\odot}$ yr$^{-1}$ \citep{Seaquist90}. Symbiotic systems have also been ruled out for the majority of SNe Ia based on their radio upper limits \citep{Horesh2012, Chomiuk2012, Torres2014, Chomiuk2016, Pellegrino2020, Lundqvist2020, Burke21}. Many models involving main-sequence companions however are still allowed within our limits of SN 2020nlb (see colored regions in Figure~\ref{fig:CSM}).

\section{X-ray Constraints on Circumstellar Material}\label{sec:x_constrain}

Deep X-ray observations can constrain the density of the circumstellar environment in the region around the supernova, which has inevitably been shaped by any mass loss in the progenitor system prior to explosion.  Within $\delta t$$\lesssim$40 days of explosion, the X-ray emission in low density environments, such as that expected from SNe Ia, is dominated by inverse Compton scattering of photospheric photons by relativistic electrons accelerated by the SN shock \citep{Chevalier06}.  A generalized formalism for the inverse Compton luminosity was developed by \citet{Margutti12}, building off of \citet{Chevalier06}, which depends on the bolometric luminosity of the SN ($L_{IC}$$\propto$$L_{bol}$), the SN ejecta mass ($M_{ej}$) and explosion energy ($E$), the density structure of the SN ejecta ($\rho_{SN}$), the density structure of the CSM ($\rho_{CSM}$), and the number of electrons and their energy distribution ($p$ and $\epsilon_e$), which is responsible for the upscattering of the optical photons to X-ray energies.   Here we adopt this formalism for determining the X-ray inverse Compton luminosity, using many of the assumptions made in that work which have been utilized to constrain the low density CSM around several other nearby SNe Ia -- most stringently for SN~2011fe \citep{Margutti12} and SN~2014J \citep{Margutti14}, but also SN~2012cg \citep{Shappee18}, SN~2018oh \citep{Shappee19}, and the SN Iax 2014dt \citep{Stauffer21}.  We also use the bolometric luminosities discussed in Section~\ref{sec:lum} for our results here. 

As in previous work, we assume the outer density of the SN scales as $\rho_{SN}$$\propto$$R^{-n}$ with n=10 \citep[see e.g.,][for compact progenitors]{Matzner99}.  The electrons are assumed to be distributed like a power-law distribution dependent on the Lorentz factor ($\gamma$), $n_{e}(\gamma)$$\propto$$\gamma^{-p}$ with index $p$=3; this value is supported by observations of SN Ib/c shocks \citep[e.g.][]{Soderberg06}.  We assume the fraction of post-shock energy density in relativistic electrons is $\epsilon_e$=0.1, as other SN shocks have indicated \citep{Chevalier06}; CSM density limits scale as ($\epsilon_e$/0.1)$^{-2}$ for any variation in this parameter \citep{Margutti12}.  Unlike similar radio constraints on the CSM, X-ray observations do not require any assumptions about magnetic field-related parameters (see dotted lines in Figure~\ref{fig:CSM}).  As both SN~2017cbv and SN~2020nlb appear to be `normal' SNe Ia we adopt an ejecta mass of $M_{ej}$=1.4 M$_{\odot}$ and explosion energy of $E$=10$^{51}$ erg for each; these are the same values adopted in the X-ray analysis for SN~2011fe and SN~2014J \citep{Margutti12,Margutti14}.

We explore two different scenarios for the CSM environment: 1) a constant-density ISM-like CSM  ($\rho_{CSM}$=constant) and 2) a wind-like CSM ($\rho_{CSM}$$\propto$$R^{-2}$).  In a simple case, a star that has been losing material at a constant rate, $\dot{M}$, with wind velocity, $v_w$, has a `wind-like' CSM with $\rho_{CSM}$ = $\dot{M}/(4\pi R^2 v_w)$, and would be a signature of the single degenerate scenario.  Double degenerate scenarios would have `clean' environments with no CSM or a low density ISM-like CSM (although see discussion below).  We do not consider asymmetric or CSM configurations with cavities explicitly here with our observational constraints, although we discuss a wide variety of CSM geometries in the following section.

For an ISM-like medium, and using the \citet{Margutti12} formalism along with the bolometric luminosities in Section~\ref{sec:lum}, we derive limits of $n_{CSM}$$<$ 35.8 cm$^{-3}$ and $<$65 cm$^{-3}$ for SN~2017cbv and SN~2020nlb, respectively.  Given the epoch that the X-ray data were taken, they probe radii of $R_{ISM}$=1.4$\times$10$^{16}$ cm and 1.2$\times$10$^{16}$ cm, respectively, for SN~2017cbv and SN~2020nlb.  These contraints are within an order of magnitude of those from SN 2011fe ($<$166 cm$^{-3}$) and SN 2014J ($<$3.5 cm$^{-3}$).

Similarly, for a wind-like medium and fiducial wind velocity of $v_w$=100 km s$^{-1}$, we constrain the progenitor mass loss rate to be $\dot{M}$$<$7.2$\times$10$^{-9}$ M$_{\odot}$ yr$^{-1}$ and $\dot{M}$$<$9.7$\times$10$^{-9}$ M$_{\odot}$ yr$^{-1}$ for SN 2017cbv and SN~2020nlb, respectively.  At the epoch of the {\it Chandra} observations, these correspond to radii of 1.9$\times$10$^{16}$ cm and 1.6$\times$10$^{16}$ cm for the two SNe.  These constraints are weaker than those of SN~2011fe ($\dot{M}$$\lesssim$2$\times$10$^{-9}$ M$_{\odot}$ yr$^{-1}$) and SN~2014J ($\dot{M}$$\lesssim$1$\times$10$^{-9}$ M$_{\odot}$ yr$^{-1}$), but are the same order of magnitude.

In Figure~\ref{fig:lums} we plot the time evolution of the expected Inverse Compton X-ray emission corresponding to the limits obtained for both the ISM and wind-like CSM scenarios described above, alongside our {\it Chandra} X-ray limits. In the following section we discuss the physical implications for the SN progenitor systems that these constraints provide.

\section{Discussion} \label{sec:discussion}

We have presented deep {\it Chandra} X-ray observations of SN~2017cbv and SN~2020nlb around optical maximum light in order to constrain the environment around these SN Ia explosions.  In addition to this data, we have presented VLA radio observations and a deep nebular spectrum of SN~2020nlb to further constrain the progenitor of this nearby SN.  The X-ray observations double the sample of normal SNe Ia with such deep limits, and regardless of the assumed circumburst density profile, they imply a `clean' environment at distances of $R$$\lesssim$10$^{16}$ cm from the explosion.   We focus our discussion on the constraints that these data provide on the progenitor system in this section.

\subsection{Single Degenerate Scenarios with Near-Steady Mass Loss}\label{sec:sd}

In the single degenerate scenario, where a white dwarf near the Chandrasekhar mass is accreting from a companion star (at a rate of $\dot{M}_{acc}$$\gtrsim$3$\times$10$^{-7}$ $M_{\odot}$ yr$^{-1}$ for steady hydrogen burning; \citealt{Shen07}), material can be lost to the surrounding environment via donor star winds, non-conservative mass transfer through Roche lobe overflow or even winds from the accreting white dwarf itself -- inverse Compton X-ray emission around maximum light can probe all of these scenarios.  

Symbiotic binary systems are plausible SN Ia progenitor systems, where the white dwarf accretes wind material from an evolved giant star.  Mass-loss rates range from $\dot{M}$$\approx$5$\times$10$^{-9}$ to 5$\times$10$^{-6}$ $M_{\odot}$ yr$^{-1}$ with wind velocities $v_w$$\lesssim$100 km s$^{-1}$ \citep[e.g.][see also discussion in \citealt{Meng16}]{Seaquist90,Patat11,Chen11}, as illustrated in the appropriate block of Figure~\ref{fig:CSM}.  Our new X-ray observations (complemented by our radio observations of SN~2020nlb) can rule out all plausible symbiotic systems.  While the sample size is still small, combining the current X-ray results with those from SN~2014J and SN~2011fe suggest that symbiotic-like CSM is uncommon around SN Ia systems.  Using their much larger sample of SNe Ia with radio observations, \citet{Chomiuk2016} similarly conclude that $\lesssim$10\% of SNe Ia are in symbiotic systems.

It is also possible that lower mass, nondegenerate companions (main sequence, subgiant or helium stars) are influencing the CSM environment through non-conservative mass transfer.  Here the nondegenerate secondary star fills its Roche lobe and some material is lost to the environment at the outer Lagrange point. We estimate the typical velocity of this lost and ejected material to be a few hundred km s$^{-1}$, the same order as the orbital velocity of the white dwarf, with an upper limit of $\sim$600 km s$^{-1}$, corresponding to the limit seen in stable nuclear burning white dwarfs \citep{Deufel99}.  The fraction of material lost is unknown, but is often assumed to be $\epsilon_{\rm loss}$$\approx$1\%, thus leading to the expected mass loss range of $\dot{M}$$\approx$0.3--3$\times$10$^{-8}$ $M_{\odot}$ yr$^{-1}$ shown in Figure~\ref{fig:Xrayfig}.   
Our new observations of SN~2017cbv and SN~2020nlb cut into this region of parameter space (see Figure~\ref{fig:CSM}), but do not completely rule out this `Lagrangian Losses' scenario.  Indeed, if the real fraction of material lost at the outer Lagrange points is significantly less than $\epsilon_{\rm loss}$$\approx$1\%, than this scenario would remain plausible for all X-ray constraints  published thus far, and it is likely that a next-generation X-ray mission is necessary to fully rule out this scenario. 

If the accretion rate from a nondegenerate companion is high enough, optically thick winds from the white dwarf are expected to develop, which happens around a critical value of $\dot{M}_{\rm acc}$$\sim$7$\times$10$^{-7}$ $M_{\odot}$ yr$^{-1}$ depending on the hydrogen mass fraction and white dwarf mass \citep{Hachisu99,Han04,Shen07}, leading to the range of allowed mass losses associated with the `Accretion Winds' scenario in Figure~\ref{fig:CSM} (although see the recent models of \citealt{Dragulin16} which suggest a low density cavity near the progenitor system which would lead to weaker limits).  The associated white dwarf wind velocities can be up to a few thousand km s$^{-1}$ \citep{Hachisu99}, similar to the outflows seen in X-ray luminous, nuclear burning white dwarfs \citep{Cowley98}.  Our X-ray constraints on SN~2017cbv and SN~2020nlb largely rule out this single degenerate progenitor scenario (as have the X-ray observations of SN~2011fe and SN~2014J), although a small region of the allowed parameter space (at high velocities, $\gtrsim$1000 km s$^{-1}$, and low mass loss rates, $\sim$10$^{-7}$ $M_{\odot}$ yr$^{-1}$) are still viable.  

\subsection{Non-steady mass loss in the single degenerate scenario}

Our X-ray analysis is most sensitive to progenitor scenarios with continuous mass loss up until the point of the SN explosion, although there are many instances where this may not be the case, even within the single degenerate scenario. If there is non-continuous mass loss in the time period just preceding explosion, our observations would be insensitive to any material inside or outside of  $R$$\approx$10$^{16}$ cm.

Recurrent nova systems are plausible SN Ia progenitors, with a high mass white dwarf ($>$1.3 M$_{\odot}$) accreting at 10$^{-8}$--10$^{-7}$ M$_{\odot}$ yr$^{-1}$ \citep{Livio92,Yaron05}, and experiencing repeated nova explosions due to unsteady hydrogen burning.  The recurrent nova system RS Oph is a prominent example of this class of objects \citep{Patat11}, where the white dwarf increases in mass with time \citep{Hachisu01} unlike in classical nova systems.  The SN Ia PTF11kx, for instance, showed clear evidence of nova shells in multi-epoch high resolution spectroscopy and is suggested to have a recurrent nova progenitor system \citep{Dilday12}, as have other SNe Ia with signs of interaction \citep[e.g. SN~2002ic;][]{WoodVasey06}.  

The CSM for recurrent nova systems is expected to be complex, and may include relatively high density nova shells, with nearly evacuated regions or wind material from the donating star in between; it is also possible there is a central cavity with no material as well \citep[see discussion in ][]{Dimitriadis14,Darnley19}.  The exact geometry of the CSM will depend on the time since the last nova outburst, and the outbursting history.
In this scenario, we would need to be very lucky to perform our X-ray observations at the time when the blast wave is interacting with a nova shell, as in between the CSM density would be $<$10$^{-2}$ cm$^{-3}$ \citep[e.g., see Figure 9 in ][]{Dimitriadis14}, far below our constraints of $\lesssim$36--65 cm$^{-3}$. In any case, neither SN~2017cbv or SN~2020nlb show any signs of interaction in their spectra \citep[see the high resolution optical spectroscopic sequence of ][in particular]{Ferretti17}.

Other scenarios where the mass loss stops prior to explosion would also yield non-detections in our X-ray data if the material is all at $\gtrsim$10$^{16}$ cm, or if the delay between the cessation of mass loss and explosion were $t$$\gtrsim$30$\times$(v$_w$/100 km s$^{-1}$)$^{-1}$ yr.  One model where this can occur is the so-called spin-up/spin-down scenario, where the accreting white dwarf gains enough angular momentum during the accretion process that it will not explode at the Chandrasekhar mass, and must spin down before it explodes \citep{distefano11,Justham11}.  Another scenario with potentially long delays between mass loss and the explosion is the core degenerate scenario, where the white dwarf merges with the core of an asymptotic giant branch star at the end of the common envelope phase \citep{Kashi11,Ilkov12}; here again, the rapid rotation of the white dwarf may prevent it from exploding immediately.  In both cases, the key unknown is the delay time between the completion of mass transfer and explosion, and is difficult to assess because the physical mechanism of the spin-down is not certain although calculations have suggested delays ranging from $\lesssim$10$^5$ yr to $\gtrsim$10$^9$ yr \citep{Lindblom99,Yoon05,Ilkov12,Hachisu12}; if one of these single degenerate scenarios were responsible for a sizable proportion of SNe Ia, it would be difficult to discern with X-ray observations.

\subsection{White dwarf -- white dwarf progenitors}

In the double degenerate scenario, two white dwarfs coalesce and lead to the final SN Ia explosion.  The general prediction for this scenario is for a `clean' circumbinary environment on scales greater than $R$$\gtrsim$10$^{14}$ cm \citep{Fryer10,Shen12,Raskin13}, similar to the ISM.  Despite this, there are several ways in which a white dwarf -- white dwarf merger can enrich the circumbinary environment. 
We briefly discuss these scenarios in the context of our X-ray limits at $\sim$10$^{16}$ cm.

One feature of white dwarf--white dwarf merger calculations is the tidal stripping and ejection of mass just prior to coalescence, consisting of 10$^{-4}$--10$^{-2}$ M$_{\odot}$ of material moving at $\sim$2000 km s$^{-1}$ in the equatorial region \citep{Dan11,Raskin13}, equivalent to an effective mass loss rate of $\dot{M}$$\approx$10$^{-2}$--10$^{-5}$ $M_{\odot}$ yr$^{-1}$ \citep{Raskin13}.  If this CSM, produced during the coalescence, is at appropriate radii at explosion ($\sim$10$^{16}$ cm) then it could lead to detectable inverse Compton emission in the X-ray regime.  The key parameter is the time between coalescence (and tidal tail ejection), and the SN explosion, which needs to be $\sim$10$^{8}$~s to reach the radii that X-ray observations are sensitive to.  Our observations rule out such delays between coalescence and explosion, although we cannot comment on explosions that occur on a dynamical ($\sim$10$^{2}$--10$^{3}$ s) or viscous timescale ($\sim$10$^{4}$--10$^{8}$ s); longer timescales ($\gg$10$^{8}$ s) are also not excluded by our data.

One variant of the double degenerate scenario, the `double detonation' model, posits a progenitor configuration with a carbon oxygen white dwarf accreting from a helium white dwarf companion \citep[e.g.][]{Nomoto82, Livne90,Woosley94}.  In certain scenarios, hydrogen rich material on the surface of the helium white dwarf accretes onto the carbon-oxygen white dwarf until  convective hydrogen burning is ignited (akin to a classical nova), causing the envelope on the carbon-oxygen white dwarf to expand and overflow its Roche radius, ejecting material at $\sim$1500 km s$^{-1}$ \citep{Shen13}, typically $\approx$3--6$\times$10$^{-5}$ M$_{\odot}$.  
Unfortunately, such a shell ejection must occur $\sim$2-3 years prior to the final explosion in order for it to be detectable by our X-ray observations (assuming $\sim$1500 km s$^{-1}$ ejected velocity), which are probing $\sim$10$^{16}$ cm; the simulations presented by \citet{Shen13} indicate that the time scales for these nova-like shell ejections are more like hundreds to thousands of years prior to explosion, depending on the details of the white dwarf progenitors.  Thus, while our X-ray observations can rule out short `ejection to explosion' delay times, they are not a stringent test of this double detonation prediction.

Other double degenerate scenarios with some circumstellar material are also plausible, including mass outflows during rapid accretion events \citep[e.g.][]{Guillochon10,Dan11} and disk winds from white dwarf mergers that do not promptly detonate \citep{Ji13}, although these both require specific timing between the mass ejection and eventual explosion of roughly a few years to have material at $\sim$10$^{16}$ cm where we have X-ray constraints.  Our observations of SN~2017cbv and SN~2020nlb (along with SN~2011fe and SN~2014J) largely rule out these ejection to explosion time scales.

\section{Summary \& Future Outlook}\label{sec:summary}

In this work, we have presented deep {\it Chandra} X-ray observations of two nearby type Ia SNe around maximum light, SN~2017cbv and SN~2020nlb.  X-ray observations of SNe Ia in the time period around maximum light are sensitive to inverse Compton emission, which is caused by interaction between the SN blastwave and accelerated particles in the CSM surrounding the explosion.  As this CSM was shaped by the mass loss history of the progenitor star system leading up to the explosion, X-ray observations are a sensitive probe of the pre-explosion SN environment, and ultimately the progenitor system itself.  The analysis of deep X-ray data for two additional, normal SNe Ia doubles the sample studied in this way, adding to the ground-breaking work on SN~2011fe \citep{Margutti12} and SN~2014J \citep{Margutti14}. 

The {\it Chandra} data lead to X-ray luminosity limits of $L_X$$\lesssim$5.4$\times$10$^{37}$ and $\lesssim$5.4$\times$10$^{37}$ erg s$^{-1}$ (0.3--10 keV) at 17.9 and 15.8 days after explosion for SN~2017cbv and SN~2020nlb, respectively.  Using the inverse Compton formalism of \citet{Margutti12} \citep[see also][]{Chevalier2006}, which is appropriate for SNe Ia in low density environments in the weeks after explosion, this corresponds to n$_{CSM}$$<$36 and $<$65 cm$^{-3}$ at a radius of $R$=1.4$\times$10$^{16}$ cm and 1.2$\times$10$^{16}$ cm for the two SNe, assuming a constant density ISM-like circumstellar medium.  If we assume a wind-like medium (where $\rho_{CSM}$ = $\dot{M}/(4\pi R^2 v_w)$), we obtain mass loss limits of $\dot{M}$$<$7.2$\times$10$^{-9}$ and $<$9.7$\times$10$^{-9}$ M$_{\odot}$ yr$^{-1}$ for a fiducial wind velocity of $v_w$=100 km s$^{-1}$ at a radius of $R$=1.9$\times$10$^{16}$ cm and 1.6$\times$10$^{16}$ cm. These limits rule out several prospects for the single degenerate scenario, including all symbiotic progenitor star systems, as well as large portions of the parameter space associated with mass loss at the outer Lagrange point and white dwarf accretion winds. It is difficult for current X-ray observations to rule out double degenerate scenarios as the expectation is for a generally `clean' environment around the SN, unless the timing between the ejection of tidal tails (or other material) is properly timed with the subsequent explosion such that there is significant material at $\sim$10$^{16}$ cm (which the X-ray observations probe).

In addition to our X-ray analysis, we have also presented other complementary constraints on the progenitor system of SN~2020nlb, and complementary constraints have already been published for SN~2017cbv.  Using a nebular phase spectrum of SN~2020nlb, we set strong constraints on any swept up hydrogen or helium from a nondegenerate companion star in the single degenerate scenario, obtaining limits that were $\sim$2--3 orders of magnitude below expectations for that channel; our data would have also detected the H$\alpha$ features recently observed in three fast declining SN Ia \citep{Kollmeier19,Prieto20,Elias21}.  We also obtained VLA data of SN~2020nlb around 15 days after the explosion, and the nondetection allowed us to rule out most plausible symbiotic progenitor systems, in agreement with the X-ray data. These complementary constraints on the progenitor of SN~2020nlb stand alongside the other SN studied in X-rays in this paper, SN~2017cbv.  Although SN~2017cbv had an early, blue light curve bump that may have signalled interaction with a nondegenerate companion star \citep{Hosseinzadeh17}, neither nebular spectroscopy \citep{Sand18_2017cbv} nor the X-ray constraints in the current work yield any further hints of the single degenerate scenario.  

Future directions for X-ray constraints on the CSM around nearby SNe Ia are clear.  Even with the current study, only four nearby SNe Ia have been studied to sufficient depths to rule out some standard single degenerate scenarios which predict winds or accretion losses.  X-ray CSM constraints on a statistical sample are necessary to make population-level statements about the SN Ia progenitor system and its diversity, and it is vital to have observations spanning the range of SN Ia properties (including luminosity, light curve parameters, and sub-type).  We also advocate for multiple observational probes of the SN Ia progenitor system be made for each nearby SN Ia (e.g. high cadence early light curves to search for `bumps', late time search for narrow hydrogen and helium lines, radio CSM constraints, etc), as each technique has its own strengths, dependence on modeling and systematic uncertainties.  By combining probes for the few SNe Ia that are near enough, progress can be made.

\acknowledgments

The scientific results reported in this article are based to a significant degree on observations made by the Chandra X-ray Observatory (Obs ID: 20055, 23314, 23315.  We thank H. Tananbaum and the entire {\it Chandra} team for making the X-ray observations possible.

Support for this work was provided by the National Aeronautics and Space Administration through Chandra Award Number DDO-21119X issued by the Chandra X-ray Center, which is operated by the Smithsonian Astrophysical Observatory for and on behalf of the National Aeronautics Space Administration under contract NAS8-03060.  This work makes use of observations from the Las Cumbres Observatory network.  The LCO team is supported by NSF grants AST-1911225 and AST-1911151 and NASA Swift grant 80NSSC19K1639.  Research by SV  is supported by NSF grants AST–1813176 and AST- 2008108.  Time domain research by D.J.S.\ is also supported by NSF grants AST-1821987, 1813466, 1908972, \& 2108032, and by the Heising-Simons Foundation under grant \#2020-1864. S.K.S.\ and L.C.\ are supported by NSF grant NSF AST-1907790 and the Packard Foundation. L.G.\ acknowledges financial support from the Spanish Ministry of Science, Innovation and Universities (MICIU) under the 2019 Ram\'on y Cajal program RYC2019-027683 and from the Spanish MICIU project PID2020-115253GA-I00. LW is sponsored (in part) by the Chinese Academy of Sciences (CAS) through a grant to the CAS South America Center for Astronomy (CASSACA) in Santiago, Chile.

Observations reported here were obtained at the MMT Observatory, a joint facility of the University of Arizona and the Smithsonian Institution. The National Radio Astronomy Observatory is a facility of the National Science Foundation operated under cooperative agreement by Associated Universities, Inc.

We are grateful to C. Burns for his help with the SNooPy software package.


\vspace{5mm}
\facilities{CXO (ACIS), Las Cumbres Observatory (Sinistro), FTN (FLOYDS), MMT (Blue Channel spectrograph), Swift (UVOT), VLA}

\software{  astropy \citep{2013A&A...558A..33A,astropy}, 
The IDL Astronomy User's Library \citep{IDLforever}, CIAO \citep{CIAO}, Supernova IDentification software package \citep[\texttt{SNiD};][]{snid}, \texttt{Astrometrica} \citep{Raab2012}, SNooPy \citep{Burns11,Burns14}, HEASoft \citep{heasoft}, \texttt{lcogtsnpipe} \citep{Valenti16}, \texttt{DoPHOT} \citep{Schechter93}
          }

\bibliography{main_v2}
\bibliographystyle{aasjournal}

\end{document}